\documentclass[prb,aps,twocolumn,showpacs]{revtex4-2} 
\usepackage{amsmath,amssymb,amsthm}
\usepackage{graphicx}
\usepackage{subfigure}
\usepackage{amsmath}
\usepackage{amsfonts,amssymb}
\usepackage{bm}
\usepackage{float}
\usepackage{color}
\usepackage{sidecap}
\allowdisplaybreaks
\usepackage{soul}
\setstcolor{red}
\usepackage[normalem]{ulem}
\usepackage{hyperref}
\hypersetup{colorlinks=true,breaklinks,urlcolor=blue,linkcolor=blue,citecolor=blue}

\begin{document}

\title{Non-Hermitian skin effect edge}
\author{Qi-Bo Zeng}
\email{zengqibo@cnu.edu.cn}
\affiliation{Department of Physics, Capital Normal University, Beijing 100048, China}

\begin{abstract}
Non-Hermitian skin effect (NHSE) is a unique phenomenon studied intensively in non-Hermitian systems during the past few years. In this work, we discuss the energy dependence of NHSE by introducing nonreciprocity beyond the nearest-neighboring hopping in the one-dimensional lattices. The direction of NHSE reverses as the eigenenergy of the system under open boundary conditions (OBCs) sweeps across some critical energies. To characterize such a phenomenon, we introduce the concept of non-Hermitian skin effect edges, which separate the eigenstates localized at opposite ends of the lattice in the OBC spectrum. We find the skin effect edges are determined by the self-intersections in the spectrum under periodic boundary conditions (PBCs), which are topological as the winding number of the PBC spectrum changes sign when crossing them. Moreover, the NHSE edges will disappear when the self-intersections merge into a single point by tuning the system parameters. Our work reveals the intricate interplay between NHSE and nonreciprocal hopping in non-Hermitian systems.  
\end{abstract}
\maketitle
\date{today}

\section{Introduction}
Non-Hermitian physics has attracted intensive attention recently~\cite{Cao2015RMP,Konotop2016RMP,Ganainy2018NatPhy,Ashida2020AiP,Bergholtz2021RMP}. It is known that non-Hermitian Hamiltonians can describe open systems effectively and have been exploited to study the properties of various classical~\cite{Makris2008PRL,Klaiman2008PRL,Guo2009PRL,Ruter2010NatPhys,Lin2011PRL,Regensburger2012Nat,Feng2013NatMat,Peng2014NatPhys,Wiersig2014PRL,Hodaei2017Nat,Chen2017Nat} as well as quantum systems~\cite{Brody2012PRL,Lee2014PRX,Li2019NatCom,Kawabata2017PRL,Hamazaki2019PRL,Xiao2019PRL,Wu2019Science,Yamamoto2019PRL,Yamamoto2019PRL,Naghiloo2019NatPhys,Matsumoto2020PRL}. Though the energy spectra of non-Hermitian Hamiltonians are in general complex, the discovery of real spectra in systems with $\mathcal{PT}$-symmetry~\cite{Bender1998PRL,Bender2002PRL,Bender2007RPP} or pseudo-Hermiticity~\cite{Mostafazadeh2002JMP,Mostafazadeh2010IJMMP,Moiseyev2011Book,Zeng2020PRB1,Kawabata2020PRR,Zeng2021NJP} has attracted a persistent exploration of non-Hermitian systems in the past two decades. 

Another unique phenomenon in non-Hermitian systems is the non-Hermitian skin effect (NHSE), where a macroscopic number of eigenstates accumulate at a boundary of the system~\cite{Yao2018PRL1,Yao2018PRL2}. Recently, a large volume of research has been dedicated to investigating the exotic properties due to NHSE~\cite{Alvarez2018PRB,Alvarez2018EPJ,Lee2019PRB,Zhou2019PRB,Kawabata2019PRX,Song2019PRL,Okuma2020PRB,Xiao2020NatPhys,Yoshida2020PRR,Longhi2019PRR,Yi2020PRL,Claes2021PRB,Haga2021PRL,Zeng2022PRA,Zeng2022PRB}. For instance, it has been shown that NHSE can modify the band topology significantly and break down the conventional principle of bulk-boundary correspondence in topological phases~\cite{Yao2018PRL1,Yao2018PRL2,Kunst2018PRL,Jin2019PRB,Yokomizo2019PRL,Herviou2019PRA,Zeng2020PRB,Borgnia2020PRL,Yang2020PRL2,Zirnstein2021PRL,Zhang2022arxiv}. The spectra of such systems are very sensitive to boundary conditions~\cite{Xiong2018JPC}, which is proposed for designing new types of quantum sensors~\cite{Budich2020PRL,Koch2022PRR}. In addition, the NHSE also significantly influences the phenomenon of Anderson localization~\cite{Hatano1996PRL,Shnerb1998PRL,Gong2018PRX,Jiang2019PRB,Zeng2020PRR,Liu2021PRB1,Liu2021PRB2}. The eigenenergy spectra of the extended and localized states exhibit different topology under periodic boundary conditions (PBCs)~\cite{Zeng2020PRR}. In addition, the NHSE can induce mobility edges in the disordered non-Hermitian system, which separates the localized and extended states in the spectrum~\cite{Gong2018PRX}. It is interesting to ask whether the NHSE will also show similar energy-dependent features. Since most studies on NHSE to date have focused on systems with only nearest-neighboring (NN) hopping and the NHSE is energy independent, we need to study how the NHSE behaves in systems with nonreciprocal hopping beyond the NN sites.

To answer the above questions, we study the one-dimensional (1D) lattices with nonreciprocal hopping ranging up to the nearest-neighboring $r_d$ sites in this paper. We show that the NHSE can be energy dependent in that some eigenstates are localized at the left end of the lattice while others are localized at the right end under open boundary conditions (OBCs). The direction of the NHSE reverses as the energy sweeps across the critical energies. We thus define the concept of non-Hermitian skin effect edge, which separates the states localized at opposite ends of the 1D lattices in the energy spectrum. More interestingly, we find that the skin effect edges are determined by the self-intersections in the spectrum under PBCs. The winding number of the PBC spectrum changes sign by crossing the self-intersecting points, implying the topological feature of the skin effect edge. The self-intersections will merge into a single point as we tune the system parameters, leading to the disappearance of NHSE edges. Our work unveils the exotic properties of NHSE in non-Hermitian systems with nonreciprocal hopping.

\begin{figure}[t]
	\includegraphics[width=3.3in]{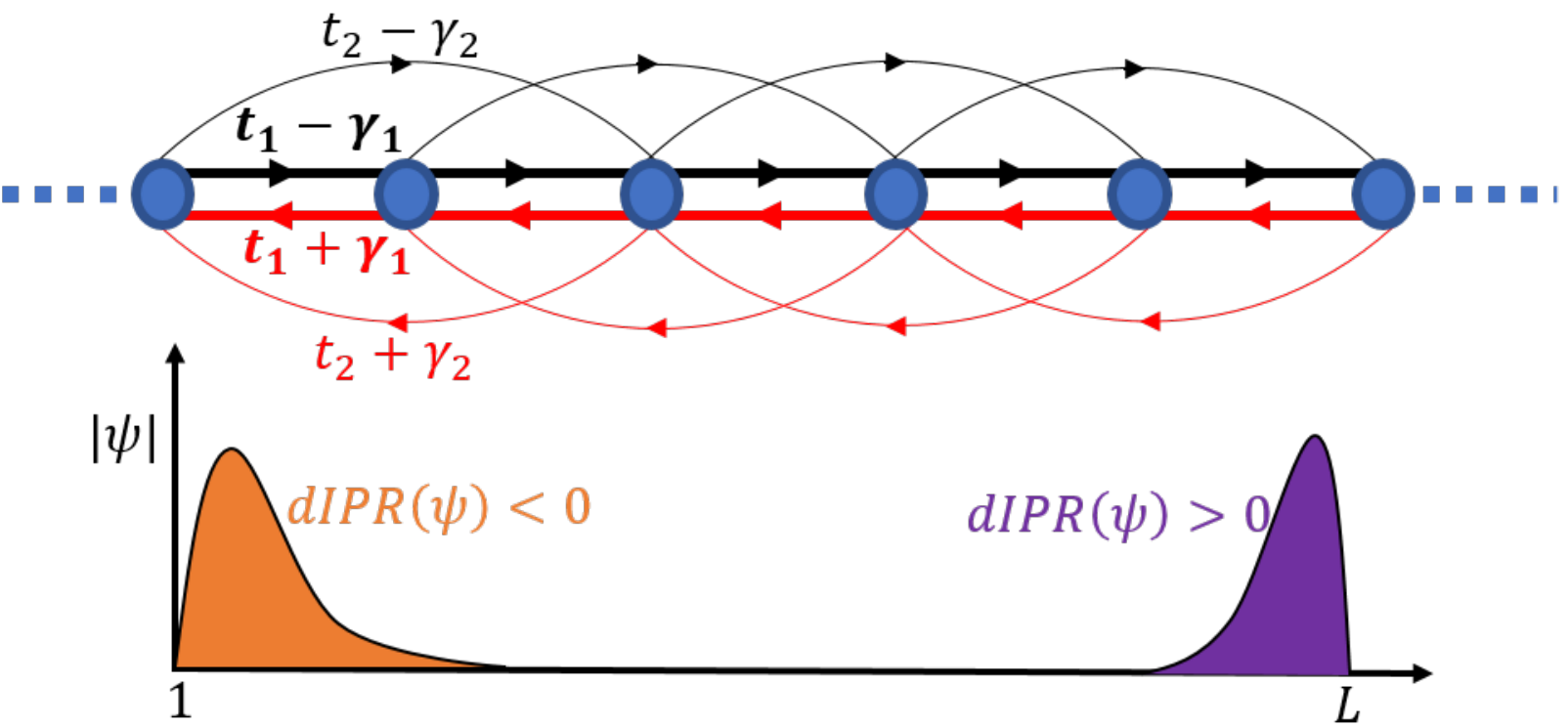}
	\caption{(Color online) Schematic of 1D lattice with nonreciprocal hoppings between the nearest- and next-nearest-neighboring sites, which are represented by $(t_1\pm \gamma_1)$ and $(t_2\pm \gamma_2)$, respectively. The lower panel shows the distribution of the eigenstates with positive and negative dIPR values.}
	\label{fig1}
\end{figure}

The rest of the paper is organized as follows. In Sec.~\ref{Sec2}, we will first introduce the general model Hamiltonian of the 1D lattices with nonreciprocal hopping up to the nearest-neighboring $r_d$ lattice sites. In Sec.~\ref{Sec3}, we introduce the NHSE edge and discuss its property in the 1D lattices with next-nearest-neighboring nonreciprocal hopping. Then we further explore the skin effect edges in systems with nonreciprocal hopping up to the third- and fourth-nearest-neighboring sites in Sec.~\ref{Sec4}. The last section (Sec.~\ref{Sec5}) is dedicated to a summary.

\section{Model Hamiltonian}\label{Sec2}
We introduce the general 1D lattices with nonreciprocal hopping described by the following model Hamiltonian
\begin{equation}\label{H}
	H = \sum_{j=1}^{L-r_d} \sum_{s=1}^{r_d} \left[(t_s - \gamma_s) c_{j+s}^\dagger c_j + (t_s + \gamma_s) c_{j}^\dagger c_{j+s} \right],
\end{equation}
where $c_j$ ($c_j^\dagger$) is the annihilation (creation) operator of spinless fermions at the $j$th site. The forward and backward hopping amplitudes are ($t_s-\gamma_s$) and ($t_s+\gamma_s$) with $t_s$ and $\gamma_s$ being real numbers. $L$ is the lattice size. $r_d$ is a positive integer and is the cutting range of the hopping, which means that the hopping extends to the nearest $r_d$ sites. If $r_d=1$, the system reduces to the Hatano-Nelson model with NN hopping. If $r_d=2$, the Hamiltonian includes both NN hopping and the next-nearest-neighboring (NNN) hopping, as shown by the schematic in the upper panel in Fig.~\ref{fig1}. It is called the $t_1$-$t_2$ model in the Hermitian case when introduced to quasiperiodic onsite potentials, which is useful in studying the mobility edge and localization properties in 1D lattices~\cite{Biddle2011PRB}. Throughout this paper, we will take $t_1=1$ as the energy unit.

With the presence of nonreciprocity in the hopping terms, there will be NHSE in the system and the eigenstates will become localized at the end of the 1D lattice. To characterize the localization of the states we may use the inverse participation ratio (IPR), which is defined as $\text{IPR}=\sum_{j=1}^L {|\Psi_{n,j}|^4}/(\langle \Psi_n | \Psi_n \rangle)^2$ and has been extensively used in describing the Anderson localization phenomenon. Here $\Psi_n$ is the right eigenstate with component $\Psi_{n,j}$ and satisfies the Schr\"odinger equation $H \Psi_n = E_n \Psi_n$, with $E_n$ being the $n$th eigenenergy. The IPR is close to 0 for the extended state but of the order $O(1)$ for the localized state. To distinguish the states localized at different ends of the 1D lattice due to NHSE, we introduce the directional IPR (dIPR) as~\cite{Zeng2022PRB}
\begin{equation}
	\text{dIPR} (\Psi_n) = \mathcal{P}(\Psi_n) \sum_{j=1}^L \frac{|\Psi_{n,j}|^4}{(\langle \Psi_n | \Psi_n \rangle)^2}, 
\end{equation}
with $\mathcal{P}(\Psi_n)=sgn \left[ \sum_{j=1}^L \left(  j- \frac{L}{2} - \delta \right) |\Psi_{n,j}| \right]$. Here $\delta \in (0,0.5)$ is a constant. $sgn(x)$ takes the sign of the argument, which is positive (negative) for $x>0$ ($x<0$). The dIPR is positive when $\Psi_n$ is localized at the right end but is negative when $\Psi_n$ is localized at the left end (see the illustration in the lower panel in Fig.~\ref{fig1}). If $\text{dIPR} \rightarrow 0$, then the state is extended.

\begin{figure}[t]
	\includegraphics[width=3.3in]{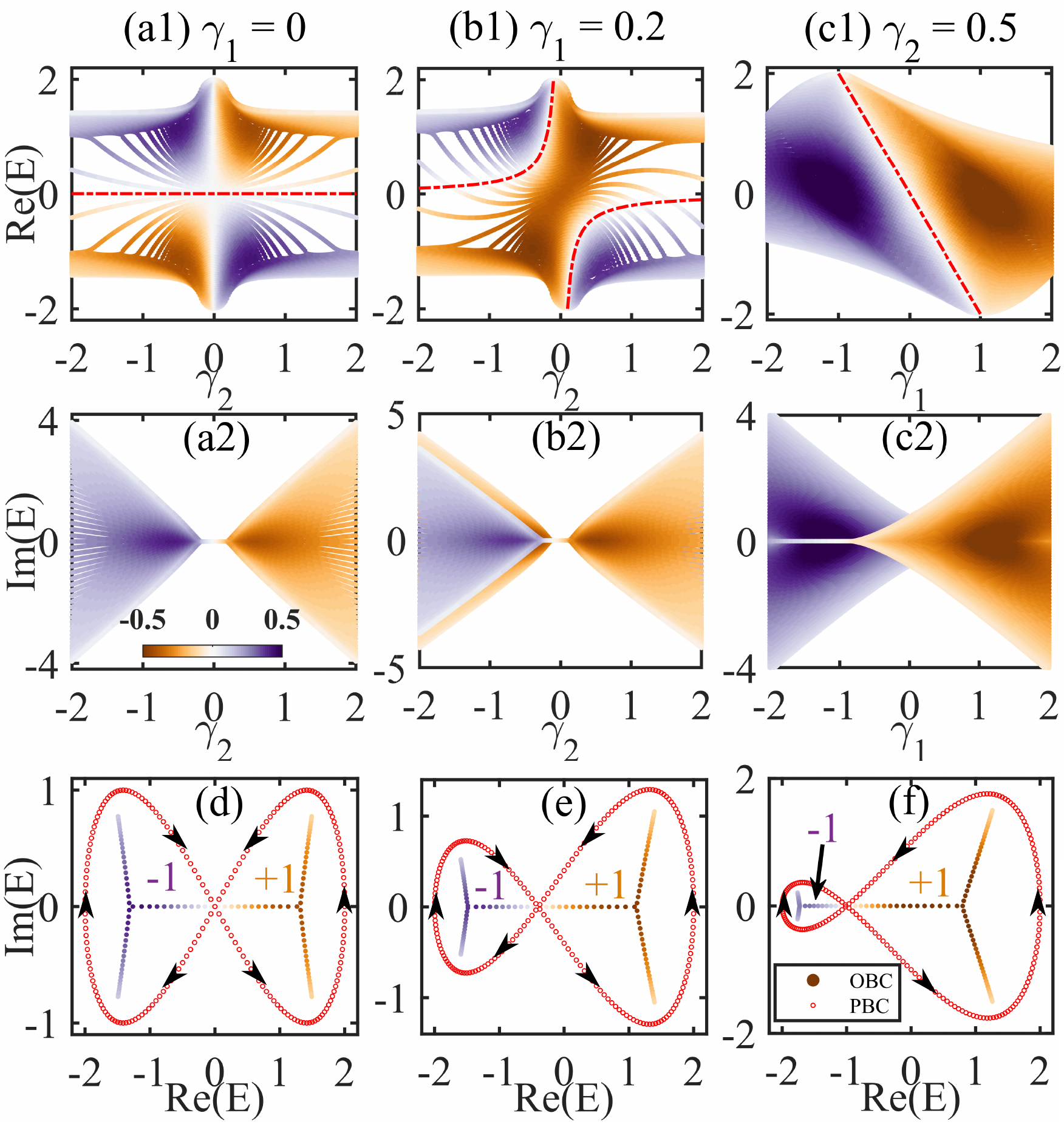}
	\caption{(Color online) (a1)-(c1) Real and (a2)-(c2) imaginary parts of the OBC eigenenergy spectra of $H_2$ with $t_2=0$ as a function of $\gamma_2$ or $\gamma_1$. The colorbar indicates the dIPR value of the eigenstate. The red dot-dashed lines are the non-Hermitian skin effect edges. The lowest panel shows the OBC (solid dots) and PBC (red empty dots) spectra of the systems with $\gamma_2=0.5$ and $\gamma_1=0.0$, $0.2$, and $0.5$ in (d)-(f), respectively. The black arrows indicate the variation of the PBC spectrum as $k$ varies. The numbers $\pm 1$ are the winding numbers for the PBC spectra. The lattice size is $L=120$.}
	\label{fig2}
\end{figure}

\section{Non-Hermitian skin effect edge}\label{Sec3}
We mainly consider the 1D lattices with NNN hopping (see Fig.~\ref{fig1}), i.e., we set $r_d=2$ in Eq.~(\ref{H}) and obtain
\begin{equation}\label{H2}
	\begin{split}
		H_2 =& \sum_{j=1}^{L-2} \left[ (t_1-\gamma_1)c_{j+1}^\dagger c_{j} + (t_2-\gamma_2) c_{j+2}^\dagger c_{j} \right. \\
		& \left. + (t_1+\gamma_1) c_{j}^\dagger c_{j+1} + (t_2+\gamma_2) c_{j}^\dagger c_{j+2} \right].
	\end{split}
\end{equation}
Notice that if the NNN hopping $t_2=\gamma_2=0$, the model reduces to the well-known Hatano-Nelson model, where all the eigenstates are localized at the same end of the lattice, i.e., the direction of NHSE is energy-independent. If we further set $t_1=0$, then there will be no skin effect in the system since the forward and backward hopping amplitudes are $|\gamma_1|$, though their signs are opposite. Thus, in order to explore the possible energy-dependence of the NHSE, we need to check the system with nonzero NNN hopping.

By transferring $H_2$ into the momentum space with $k\in [-\pi, \pi)$, we get the PBC spectrum as
\begin{equation}\label{E2k}
	E_2(k) = 2\left[t_1 \cos(k) + t_2 \cos(2k) \right] + 2i \left[\gamma_1 \sin(k) + \gamma_2 \sin(2k) \right].
\end{equation} 
To illustrate the effect of the NNN hopping on the NHSE, we first set $\gamma_1=t_2=0$ and check the properties of the following simplified Hamiltonian
\begin{equation}\label{H2'}
	H_2^\prime = \sum_{j=1}^{L-2} \left[ t_1c_{j+1}^\dagger c_{j} + t_1 c_{j}^\dagger c_{j+1} -\gamma_2 c_{j+2}^\dagger c_{j} + \gamma_2 c_{j}^\dagger c_{j+2}  \right].
\end{equation}
Due to the opposite signs of the NNN forward and backward hopping terms in the system, the skin effect shows up even though the strengths are both $|\gamma_2|$, as shown in Fig.~\ref{fig2}(a). More interestingly, when $\gamma_2>0$ (or $\gamma_2<0$), half of the eigenstates with $Re(E)<0$ [or $Re(E)>0$] are localized at the right end of the lattice; while the other half with $Re(E)>0$ [or $Re(E)<0$] are localized at the left end. This is totally different from the NHSE shown in systems with NN asymmetric hopping, where all the eigenstates localize at the same end. Here the skin effect is energy-dependent and the line $Re(E)=0$ separates the states localized at opposite ends. The direction of NHSE reverses when the energy sweeps across the line. Analogous to the definition of mobility edge, which separates extended states from localized states in disordered systems, here we define non-Hermitian skin effect edge as the energy separating the states localized at different ends of the 1D lattice. So, for the simple model in Eq.~(\ref{H2'}), the skin effect edge is the zero energy. If we have $\gamma_1 \neq 0$, the skin effect edge will deviate from zero energy, as shown in Fig.~\ref{fig2}(b). The red dot-dashed lines in the real part are the NHSE edges described by the equation $Re(E)=-\frac{\gamma_1}{\gamma_2} t_1$. If we plot the eigenenergy as a function of $\gamma_2$, the skin effect edge will be a straight line [see Fig.~\ref{fig2}(c)]. We find that for the systems with $t_2=0$, the NHSE edge is always real.

\begin{figure}[t]
	\includegraphics[width=3.3in]{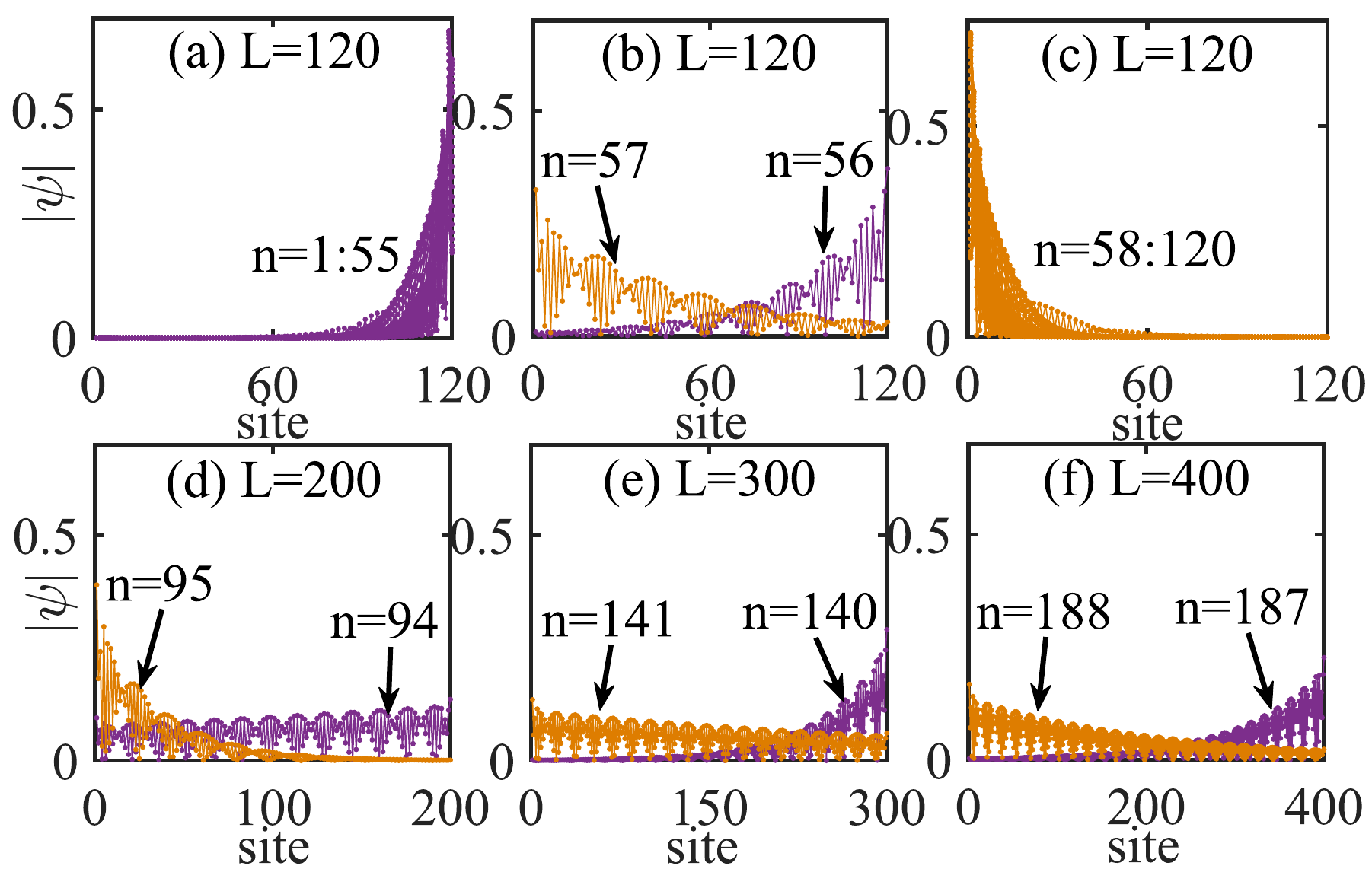}
	\caption{(Color online) Distribution of eigenstates of $H_2$ under OBC in the system with $\gamma_1=0.2$ and $\gamma_2=1$. The states are sorted according to the real parts of the eigenenergies. In (a)-(c), the lattice size is $L=120$. (b) and (d)-(f) show the eigenstates with $Re(E)$ closest to the edge at $E=-0.2$ in systems with increasing lattice sizes.}
	\label{fig3}
\end{figure}

In Fig.~\ref{fig3}, we plot the distribution of the eigenstates of the Hamiltonian $H_2$ with $t_2=0$, $\gamma_1=0.2$ and $\gamma_2=1$. The NHSE edge in this specific case is $Re(E)=-\frac{\gamma_1}{\gamma_2} t_1=-0.2$ The eigenstates are sorted according to the real parts of the eigenenergies. We can see that away from the NHSE edge, the states with $Re(E)<0.2$ are localized at the right end of the lattice, while those with $Re(E)>0.2$ are localized at the left end. For the states closest to the edge, i.e., states labeled by $n=56$ and $57$, they are not that localized, as shown in Fig.~\ref{fig3}(b). As the system size increases, there will be states with $Re(E)$ closer to the edge. The distribution of these states becomes more extended [see Fig.~\ref{fig3}(d)-\ref{fig3}(f)]. Thus if the lattice size goes to infinite, we can obtain the state corresponding to the NHSE edge, which is fully extended.

It is known that NHSE is closely connected with the point gap in the PBC spectrum~\cite{Okuma2020PRL,Zhang2020PRL}. For the energy-dependent NHSE we study here, we can also expect the existence of point gaps. In Figs.~\ref{fig2}(d)-\ref{fig2}(f), we present the OBC (solid dots) as well as PBC (red empty dots) spectra for systems with $t_1=1$, $t_2=0$, $\gamma_2=0.5$, and different $\gamma_1$ values.The PBC spectrum is inseparable and intersects with itself. The loops in the spectrum are characterized by the winding number defined as
\begin{equation}\label{W}
	W = \frac{1}{2\pi i} \int_{-\pi}^{\pi} dk \partial_k \text{arg} \left[ E(k)-E_B \right],
\end{equation}
where $E_B$ is the base energy. We find that $W$ has opposite signs for the loops enclosing states localized at the opposite ends. The energy dependence of the NHSE has also been reported in Refs.~\cite{Claes2021PRB} and \cite{Zhang2020PRL}, which are characterized by winding numbers with opposite signs. For the loop enclosing the states localized at the right (left) end, we have $W=-1$ ($W=+1$). The self-intersecting point of the PBC spectrum is the critical energy that separates the states localized at opposite ends of the system under OBC. The winding number changes sign abruptly as $E_B$ crosses the critical energy. So the NHSE edge in the energy spectrum is topological. By calculating the self-intersections $E_2^{SI}$ in the PBC spectrum, we can obtain the critical energies and thus determine the skin effect edges. To do this, suppose that we have $k_1$, $k_2 \in [-\pi, \pi)$ and $k_1 \neq k_2$, such that $E_2(k_1)=E_2(k_2)$ with $E(k)$ given in Eq.~(\ref{E2k})). For the systems with $t_2=0$, the solution to this equation gives us the NHSE edge as (see details in the Appendix)
\begin{equation}
	\text{Re}\left( E_2^{SI} \right)=-\frac{\gamma_1}{\gamma_2} t_1; \qquad \text{Im}\left( E_2^{SI} \right)=0,
\end{equation}
with the restriction $\left| \frac{\gamma_1}{2\gamma_2} \right| \leq 1$, which is consistent with the numerical results shown in Fig.~\ref{fig2}. 

\begin{figure}[t]
	\includegraphics[width=3.3in]{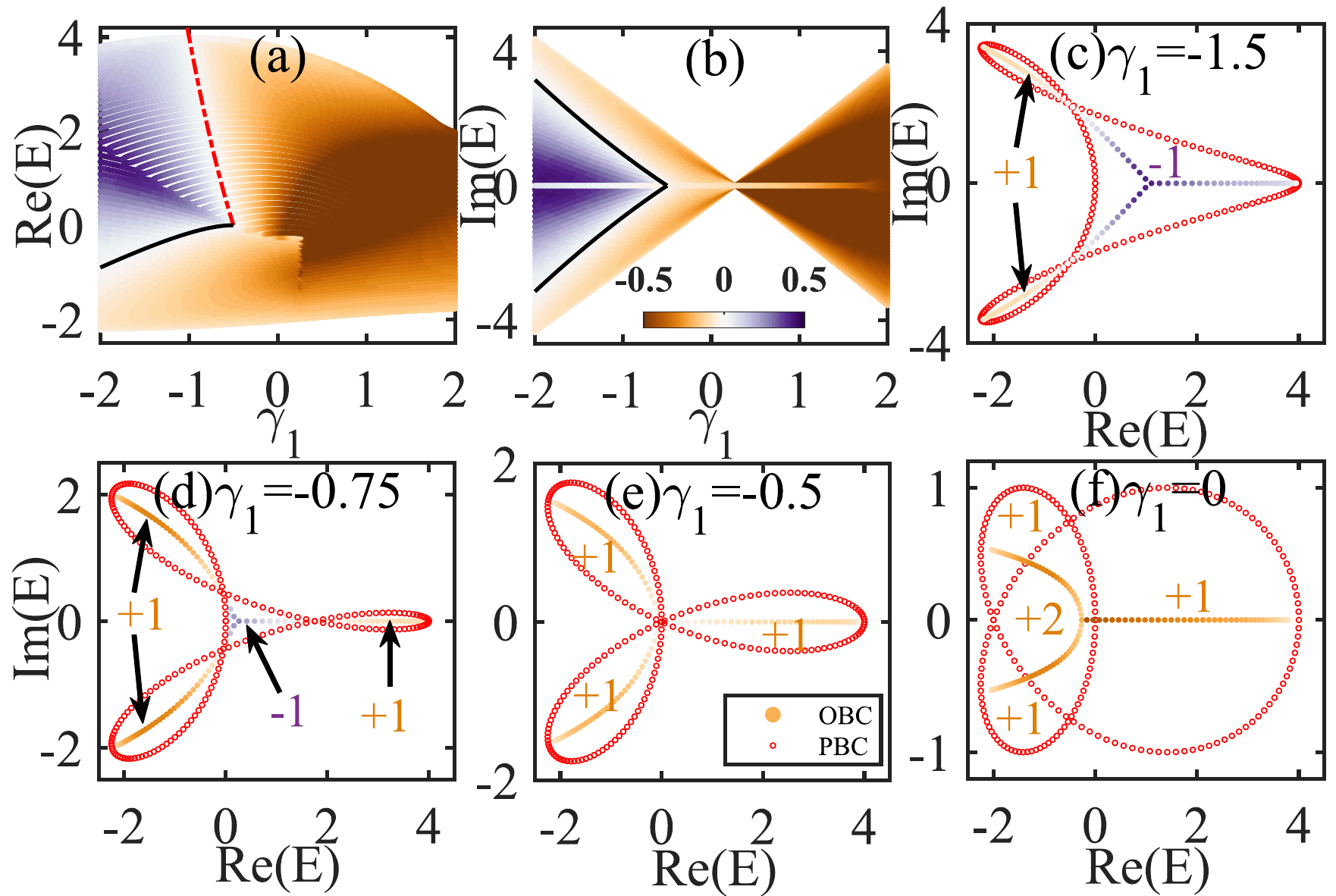}
	\caption{(Color online) (a) and (b) show the real and imaginary parts of the OBC eigenenergy spectra of $H_2$ with $t_2=1$ as a function of $\gamma_1$. (c)-(f) are the OBC (solid dots) and PBC (red empty dots) spectra with $\gamma_1$ varying from $-1.5$ to $0$. The red dot-dashed lines and black-solid lines are the skin effect edges determined by Eqs.~(\ref{SI1}) and Eq.~(\ref{SI23}). The numbers indicate the winding numbers of the loops. Here we set $\gamma_2=0.5$ and the lattice size $L=120$.}
	\label{fig4}
\end{figure}

Now we turn on the NNN hopping $t_2$ and check how the skin effect edge will change. In Fig.~\ref{fig4}(a) and \ref{fig4}(b), we present the real and imaginary parts of the eigenenergy of the system with $t_1=t_2=1$ and $\gamma_2=0.5$ as a function of $\gamma_1$. We find that different from the situation with $t_2=0$, where the edge is purely real, the skin effect edge here can be complex. There are three NHSE edges in the spectrum. One of them is real, as indicated by the red dot-dashed line in Fig.~\ref{fig4}(a). The other two are complex conjugated, as shown by the black-solid lines in the real and imaginary parts. To further illustrate the distinctive features of the energy spectra, we plot the OBC (solid dots) and PBC (red-empty dots) eigenenergy under different $\gamma_1$ values on the complex plane, as shown in Fig.~\ref{fig4}(c)-\ref{fig4}(f). When $\gamma_1<-0.5$, we can see that there are several loops in the PBC spectrum, which enclose the OBC spectrum with states localized at the different ends of the system. The corresponding winding numbers are $+1$ and $-1$, respectively. We can determine the NHSE edge by calculating the self-intersections $E_2^{SI}$ of the PBC spectrum, which gives us the following three analytical expressions (see Appendix):
\begin{equation}\label{SI1}
	\text{Re}\left(E_2^{SI1}\right) = -\frac{\gamma_1}{\gamma_2}t_1 + \left( \frac{\gamma_1^2}{\gamma_2^2}-2 \right) t_2; \text{Im}\left(E_2^{SI1}\right) = 0.
\end{equation}
\begin{equation}\label{SI23}
	\begin{split}
		&\text{Re}\left(E_2^{SI2}\right) = \text{Re}\left(E_2^{SI3}\right) = -\frac{t_1^2}{2t_2} + \frac{t_1\gamma_1}{2\gamma_2} - \frac{2t_2^2 \gamma_1}{t_1 \gamma_2 - t_2 \gamma_1}; \\
		&\text{Im}\left(E_2^{SI2}\right) = a \left[ \frac{\gamma_1}{2t_2 \gamma_2} - \left( \frac{t_1}{2t_2^2} - \frac{2\gamma_2}{t_1\gamma_2 - t_2\gamma_1} \right)\right], \\
		&\text{Im}\left(E_2^{SI3}\right) = a \left[ -\frac{\gamma_1}{2t_2 \gamma_2} + \left( \frac{t_1}{2t_2^2} - \frac{2\gamma_2}{t_1\gamma_2 - t_2\gamma_1} \right)\right],
	\end{split}
\end{equation}
where $a=\sqrt{t_1^2 \gamma_2^2 - 2t_1 t_2 \gamma_1 \gamma_2}$. These critical energies are exactly the skin effect-edges observed in the OBC spectra.

By tuning the system parameters, the self-intersections in the PBC spectrum will merge into one point. The critical parameters for this situation happening are obtained by solving the following equation:
\begin{equation}\label{eqSI}
	E_2^{SI1}=E_2^{SI2}=E_2^{SI3}.
\end{equation}
For instance, in the system with $t_1=t_2=1$, we find that when $\gamma_1=-\gamma_2$, the critical lines in the OBC spectrum intersect with each other, leading to the disappearance of the NHSE edges at the intersecting point (see Appendix for the computation). The loops in the corresponding PBC spectrum are merged at zero energy and form a trefoil, see Fig.~\ref{fig4}(e). The winding numbers for the three loops are $+1$, which indicates that the eigenstates under OBC are all localized at the left end of the lattice. If $\gamma_1$ further increases, there will be regions characterized by $W=+2$ in the PBC spectrum, as shown in Fig.~\ref{fig4}(f). However, the NHSE edges no longer exist. So, by varying parameters of the system, we can tune the skin effect edge in the OBC spectrum and obtain various spectral topologies in the PBC spectra. 

\begin{figure}[t]
	\includegraphics[width=3.3in]{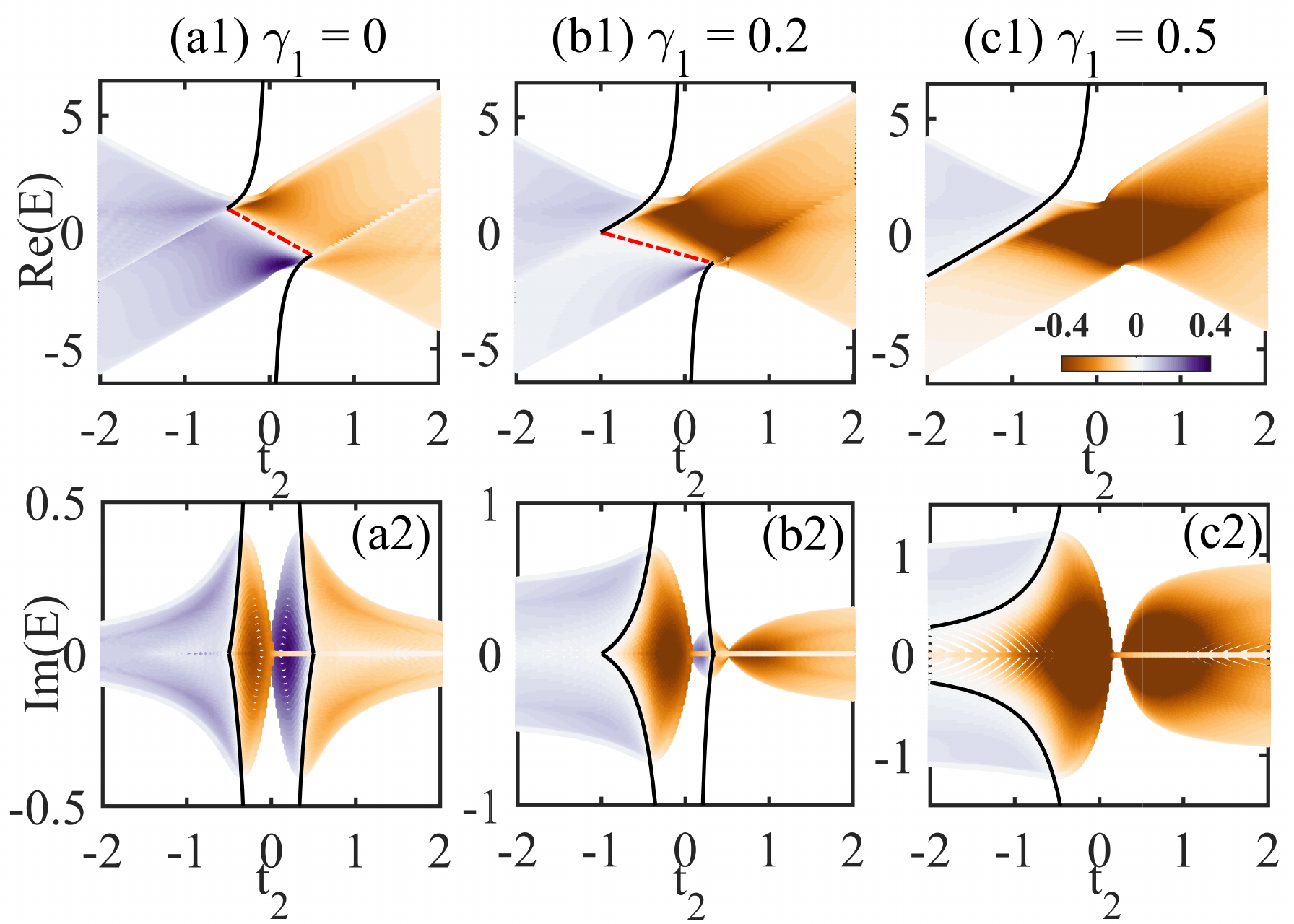}
	\caption{(Color online) Real and imaginary parts of the OBC eigenenergy of $H_2$ with $t_1=1$ and $\gamma_2=0.2$ as a function of $t_2$ when (a) $\gamma_1=0$, (b) $\gamma_1=0.2$, and (c) $\gamma_1=0.5$. The corlorbar indicates the dIPR value of the eigenstates. The red dotted lines and black solid lines are the skin effect edges in Eqs.~(\ref{SI1}) and (\ref{SI23}). The lattice size is $L=120$.}
	\label{fig5}
\end{figure}

In Fig.~\ref{fig5}, we further show the OBC spectra of the system as a function of $t_2$. The NHSE edges indicated by the lines are determined by Eq.~(\ref{SI1}) and (\ref{SI23}), respectively. As $t_2$ varies, these critical lines intersect each other, and the NHSE edges disappear. This corresponds to the merging point obtained by solving Eq.~(\ref{eqSI}). Notice that the real-valued edge (i.e., the red dot-dashed line) in Eq.~(\ref{SI1}) is restricted in the region $\left| \frac{\gamma_1}{2\gamma_2} \right| \leq 1$. So if $\gamma_1$ becomes too stronger, this NHSE edge will also disappear, as shown in Fig.~\ref{fig4}(c). 

\section{Skin effect edge in systems with $\mathbf{r_d>2}$}\label{Sec4}
The existence of skin effect edge is not limited to the model with NNN hopping. For systems with hopping beyond the next-nearest-neighboring sites, the NHSE edge still exists. For instance, we can check the system with hopping up to the third or fourth nearest-neighboring sites. The corresponding Hamiltonians are obtained by setting $r_d=3$ and $4$ in Eq.~(\ref{H}). For simplicity, we consider the systems described by $H_3^\prime$ and $H_4^\prime$, which are obtained by replacing $\gamma_2$ with $\gamma_3$ or $\gamma_4$ in Eq.~(\ref{H2'}). The model Hamiltonian with only $t_1$ and $\gamma_3$ terms is
\begin{equation}
	H_3^\prime = \sum_{j=1}^{L-3} \left[ t_1 c_{j+1}^\dagger c_{j} + t_1 c_{j}^\dagger c_{j+1} -\gamma_3 c_{j+3}^\dagger c_{j} +  \gamma_3 c_{j}^\dagger c_{j+3} \right].
\end{equation}
The model Hamiltonian with only $t_1$ and $\gamma_4$ terms is
\begin{equation}
	H_4^\prime = \sum_{j=1}^{L-4} \left[ t_1 c_{j+1}^\dagger c_{j} + t_1 c_{j}^\dagger c_{j+1} -\gamma_4 c_{j+4}^\dagger c_{j} +  \gamma_4 c_{j}^\dagger c_{j+4} \right].
\end{equation}

\begin{figure}[t]
	\includegraphics[width=3.3in]{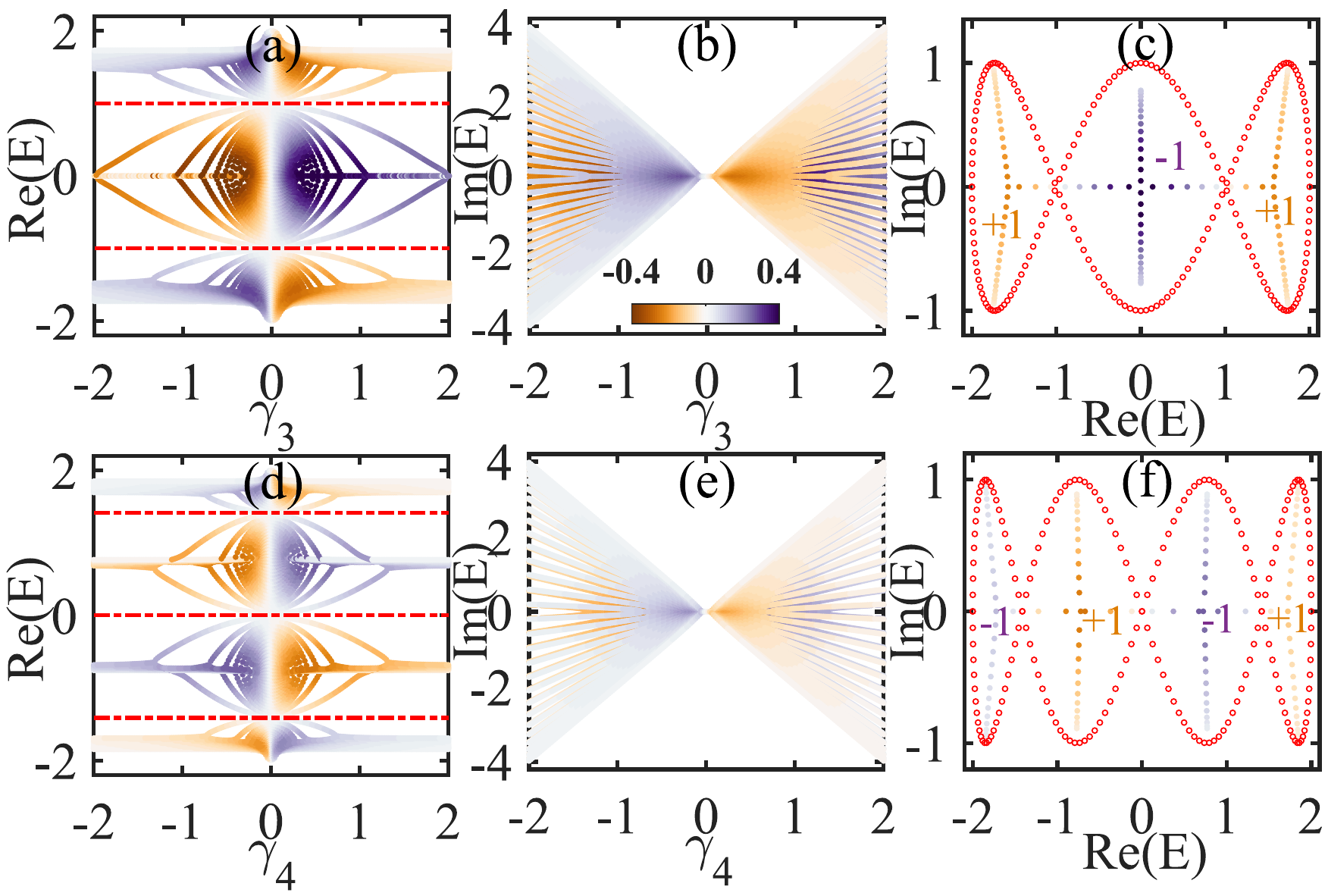}
	\caption{(Color online) The OBC eigenenergy spectra for $H_3^\prime$ and $H_4^\prime$ as a function of the nonreciprocal hopping $\gamma_3$ in (a) and (b), or $\gamma_4$ in (d) and (e). The red dot-dashed lines are the NHSE edges. (c) and (f) show the PBC (red empty dots) and OBC (solid dots) spectra when $\gamma_3, \gamma_4=0.5$, respectively. $\pm 1$ are the winding numbers of the PBC spectra. Here the lattice size is $L=120$.}
	\label{fig6}
\end{figure}

In Fig.~\ref{fig6}, we present the energy spectra of $H_3^\prime$ and $H_4^\prime$ as a function of $\gamma_3$ and $\gamma_4$. We can see that the real parts of the spectra are divided into several regions according to the positive or negative dIPR values of the eigenstates, while the imaginary parts are not. So there are NHSE edges in these systems and the critical energies are real. In Figs.~\ref{fig6}(c) and (f), we show the OBC (solid dots) and PBC (red empty dots) spectra for systems with both $\gamma_3$ and $\gamma_4$ being $0.5$ in the complex energy plane. The PBC spectrum forms loops that are characterized by $+1$ ($-1$) when the enclosed OBC eigenstates localized at the left (right) end of the lattice. The edges thus are topological as the winding number changes suddenly by crossing them. Again we can determine the NHSE edges by calculating the self-intersections in the PBC spectrum, which gives us two self-intersecting points for $H_3^\prime$ (see Appendix): $E_3^{SI1} = -E_3^{SI2} = t_1$; and three self-intersecting points for $H_4^\prime$: $E_4^{SI1} = 0, \qquad E_4^{SI2} = -E_4^{SI3} = \sqrt{2} t_1$. These are the NHSE edges indicated by the red dot-dashed lines in Figs.~\ref{fig6}(a) and \ref{fig6}(d).

\section{Summary}\label{Sec5}
In summary, we introduce the concept of non-Hermitian skin effect edge in the 1D systems with nonreciprocal hopping beyond the nearest-neighboring sites. The NHSE is energy-dependent and the edge separates the eigenstates under OBC localized at opposite ends of the system. The direction of the skin effect reverses as the energy moves across the edge. We also find that the skin effect edges are determined by the self-intersections in the PBC spectrum and are topological as the winding numbers of the PBC spectrum change sign when crossing the critical points. Our work unveils the exotic properties of the non-Hermitian systems with nonreciprocal hopping beyond the nearest-neighboring sites.

\begin{acknowledgments}
This work is supported by NSFC (Grant No. 12204326), R\&D Program of Beijing Municipal Education Commission (Grant No. KM202210028017) and Open Research Fund Program of the State Key Laboratory of Low-Dimensional Quantum Physics (Grant No. KF202109).
\end{acknowledgments}

\begin{widetext}

\section*{Appendix}
\setcounter{equation}{0}
\renewcommand{\theequation}{{A}\arabic{equation}}
In the Appendix, we provide more details on the determination of the non-Hermitian skin effect edge in the non-Hermitian lattices with nonreciprocal hopping beyond nearest-neighboring sites. We also show more numerical results of the energy spectra of the 1D lattice investigated in the main text.

\subsection{Non-Hermitian skin effect edge in systems with next-nearest-neighboring hopping}
\subsubsection{Cases with $t_2 = 0$}
To calculate the self-intersecting points in the PBC spectrum, we can suppose that we have $k_1$, $k_2 \in [-\pi, \pi)$ and $k_1 \neq k_2$, such that $E_2(k_1)=E_2(k_2)$. For the system with $t_2=0$ and $\gamma_1=0$, we have $E_2^\prime(k)=2t_1 \cos(k) + 2i \gamma_2 \sin (2k)$. Then from $E_2(k_1)=E_2(k_2)$, we have
\begin{equation}
	\begin{cases}
		t_1 \cos (k_1) = t_1 \cos (k_2) \\
		\gamma_2 \sin (2k_1) = \gamma_2 \sin (2k_2)
	\end{cases}
	\Longrightarrow 
	\begin{cases}
		\cos (k_1) = \cos (k_2) \\
		\sin (2k_1) = \sin (2k_2)
	\end{cases}
	\Longrightarrow 
	\begin{cases}
		\cos (k_1) = \cos (k_2) \\
		\sin (k_1) \cos (k_1) = \sin (k_2) \cos (k_2).
	\end{cases}
\end{equation}  
If $\cos (k_1) = \cos (k_2) \neq 0$, then we have $k_1 = -k_2$ and $\sin (k_1)=\sin (k_2)$, which further leads to $\sin (k_1) =\sin (k_2) = 0$. So we have $k_1$, $k_2 \in \left\{0, -\pi \right\}$, which cannot satisfy the condition $k_1 = -k_2$. On the other hand, if $\cos (k_1) = \cos (k_2) = 0$, then we can set $k_1=-\frac{\pi}{2}$ and $k_2=\frac{\pi}{2}$, or vice versa. Then the condition $\sin (2k_1) = \sin (2k_2)$ is also satisfied. Substituting $k_1$ or $k_2$ back into $E_2(k)$, we can obtain the self-intersecting point $E_2^{SI}$ as
\begin{equation}
	\text{Re}(E_2^{SI}) = 0, \qquad \text{Im}(E_2^{SI}) = 0.
\end{equation}
So the self-intersecting point is the zero energy point. The winding number of the PBC spectrum changes sign when it crosses this point. The states with $Re(E)<0$ and $Re(E)>0$ are localized at the opposite ends of the 1D lattices, which is consistent with the numerical results in Fig.~\ref{fig2}(a).

For the cases with $t_2 = 0$ and $\gamma_1 \neq 0$, we can determine the skin effect edge similarly. The PBC spectrum is $E_2(k)=2t_1 \cos (k) + 2i \left[\gamma_1 \sin (k) + \gamma_2 \sin (2k)\right]$. Suppose we have $k_1$, $k_2 \in [-\pi, \pi)$ and $k_1\neq k_2$, such that $E_2(k_1)=E_2(k_2)$, then we have 
\begin{equation}
	\begin{cases}
		t_1 \cos(k_1) = t_1 \cos(k_2) \\
		\gamma_1 \sin(k_1) + \gamma_2 \sin (2k_1) = \gamma_1 \sin(k_2) + \gamma_2 \sin (2k_2) 
	\end{cases}
	\Longrightarrow
	\begin{cases}
		\cos(k_1) = \cos(k_2) \rightarrow k_1 = -k_2 \\
		\gamma_1 \left[\sin(k_1) - \sin(k_2)\right] + \gamma_2 \left[\sin(2k_1)-\sin(2k_2)\right] = 0.
	\end{cases}
\end{equation}
From $k_1 = -k_2$, the second condition in the above equation becomes 
\begin{equation}
	\left[ \gamma_1 + 2\gamma_2 \cos(k_1) \right] \left[\sin(k_1) - \sin(k_2)\right] = 0.
\end{equation}
$\left[\sin(k_1) - \sin(k_2)\right] \neq 0$ due to $k_1 = -k_2$, so we have $\left[ \gamma_1 + 2\gamma_2 \cos(k_1) \right]=0$, which leads to $\cos(k_1)=-\frac{\gamma_1}{2\gamma_2}$. Substituting $\cos(k_1)$ into $E_2(k)$, we obtain the self-intersecting point as
\begin{equation}
	\text{Re}(E_2^{SI}) = -\frac{\gamma_1}{\gamma_2} t_1, \qquad \text{Im}(E_2^{SI}) = 0.
\end{equation}
Thus in the real parts of the eigenenergy spectra, the states with $Re(E)<-\frac{\gamma_1}{\gamma_2} t_1$ are localized at the right end of the lattice and those with $Re(E)>-\frac{\gamma_1}{\gamma_2} t_1$ are localized at the left end. Notice that $\cos(k_1)\leq 1$, we also have
\begin{equation}
	\left| \frac{\gamma_1}{2\gamma_2} \right| \leq 1.
\end{equation}
Thus the skin effect edge only shows up when $\left| \gamma_1 \right| \leq \left| 2\gamma_2 \right|$, which is consistent with the numerical results shown in the main text.

\subsubsection{Cases with $t_2 \neq 0$}
When $t_2 \neq 0$, the situation becomes more complicated. In Fig.~\ref{figA1}, we present the real and imaginary parts of the energy spectra as a function of $\gamma_1$ for the system with $t_1=t_2=1$ and various $\gamma_2$ values. When $\gamma_2=0$, there is only nonreciprocal hopping in the nearest-neighboring hopping amplitudes and there is no skin effect edge in the spectrum, as shown in Fig.~\ref{figA1}(a). As $\gamma_2$ becomes nonzero, the skin effect edge shows up and can be observed in both the real and imaginary parts of the spectra, which means that critical energies that separates the states localized at different ends of the lattice are not real, but complex. Moreover, there can be more than one NHSE edge in the spectra, as indicated by the red and black lines in Figs.~\ref{figA1}(b) and \ref{figA1}(c).  

\begin{figure}[t]
	\includegraphics[width=4.0in]{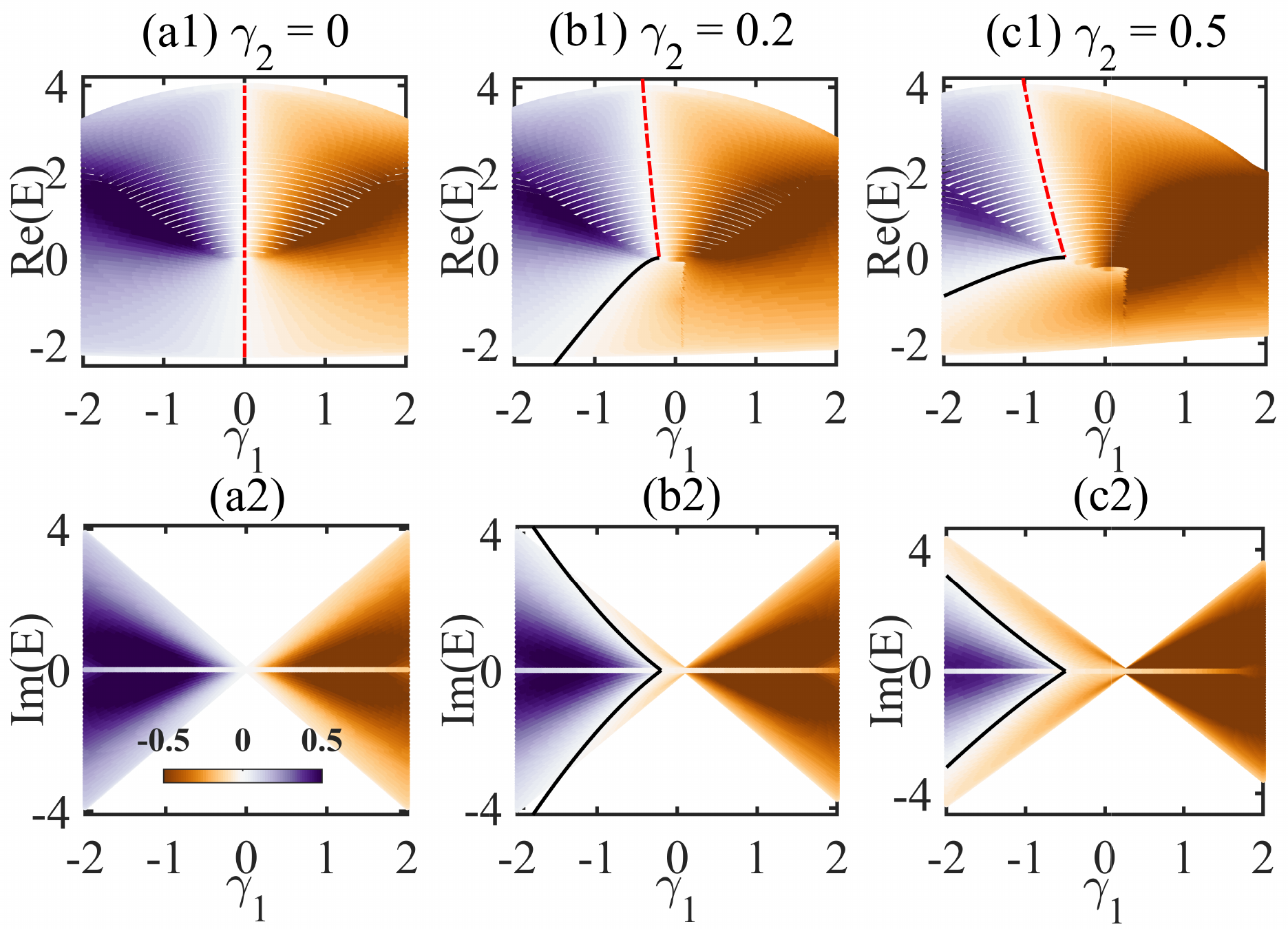}
	\caption{(Color online) Real and imaginary parts of the eigenenergy of $H_2$ with $t_1=1$ and $t_2=1$ as a function of $\gamma_1$ when (a) $\gamma_2=0$, (b) $\gamma_2=0.2$, and (c) $\gamma_2=0.5$. The corlorbar indicates the dIPR value of the eigenstates. The red dotted lines and black solid lines are the skin effect edges in Eqs.~(\ref{SI1A}) and (\ref{SI23A}). The lattice size is $L=120$.}
	\label{figA1}
\end{figure}

Next we check the connections between the OBC and PBC spectra. In Fig.~\ref{figA2}, we show both the OBC (solid dots) and PBC (red empty dots) energy spectra of systems with different $\gamma_1$ in the complex energy plane. The PBC spectra are obtained by calculating Eq.~(\ref{E2k}) for $k \in [-\pi,\pi)$. Here we have fixed the value of $\gamma_2$ at $0.5$. We find that when $\gamma_1\leq -1.0$, there are two skin effect edges in the spectra, which are the two self-intersecting points connecting the three loops in the PBC spectra. The winding number for the loops enclosing the states localized at the left and right ends of the lattice under OBC are $+1$ and $-1$, respectively, as shown in Figs.~\ref{figA2}(a) and \ref{figA2}(b). If $-1<\gamma_1<-0.5$, there are three self-intersecting points in the PBC spectrum, where two of them are complex conjugated and the remaining one is real [see Fig.~\ref{figA2}(c)]. There are four loops in the PBC spectrum. Three loops are characterized by the winding number $W=+1$, where the OBC eigenstates enclosed by these loops are localized at the left end of the lattice due to NHSE. The winding number for the remaining one loop is $-1$, and the corresponding OBC eigenstates are localized at the right end of the lattice. When $\gamma_1=-0.5$, the three loops are connected at one single point and the spectrum exhibits a trefoil structure. The winding numbers are the same for these loops, i.e., $W=+1$. There is no skin effect edge and all the states under OBC are localized at the left end [see Fig.~\ref{figA2}(d)]. As $\gamma_1$ increases, we find there are regions with winding number $W=+2$ in the PBC spectrum. However, the NHSE edges no longer exist. All the winding numbers are positive and all the eigenstates under OBC are localized at the left end, as indicated by the dIPR value shown in Figs.~\ref{figA2}(e) and \ref{figA2}(f). Such spectral structures have also been reported in Ref.~\cite{Zeng2022PRA}.
\begin{figure}[t]
	\includegraphics[width=4.0in]{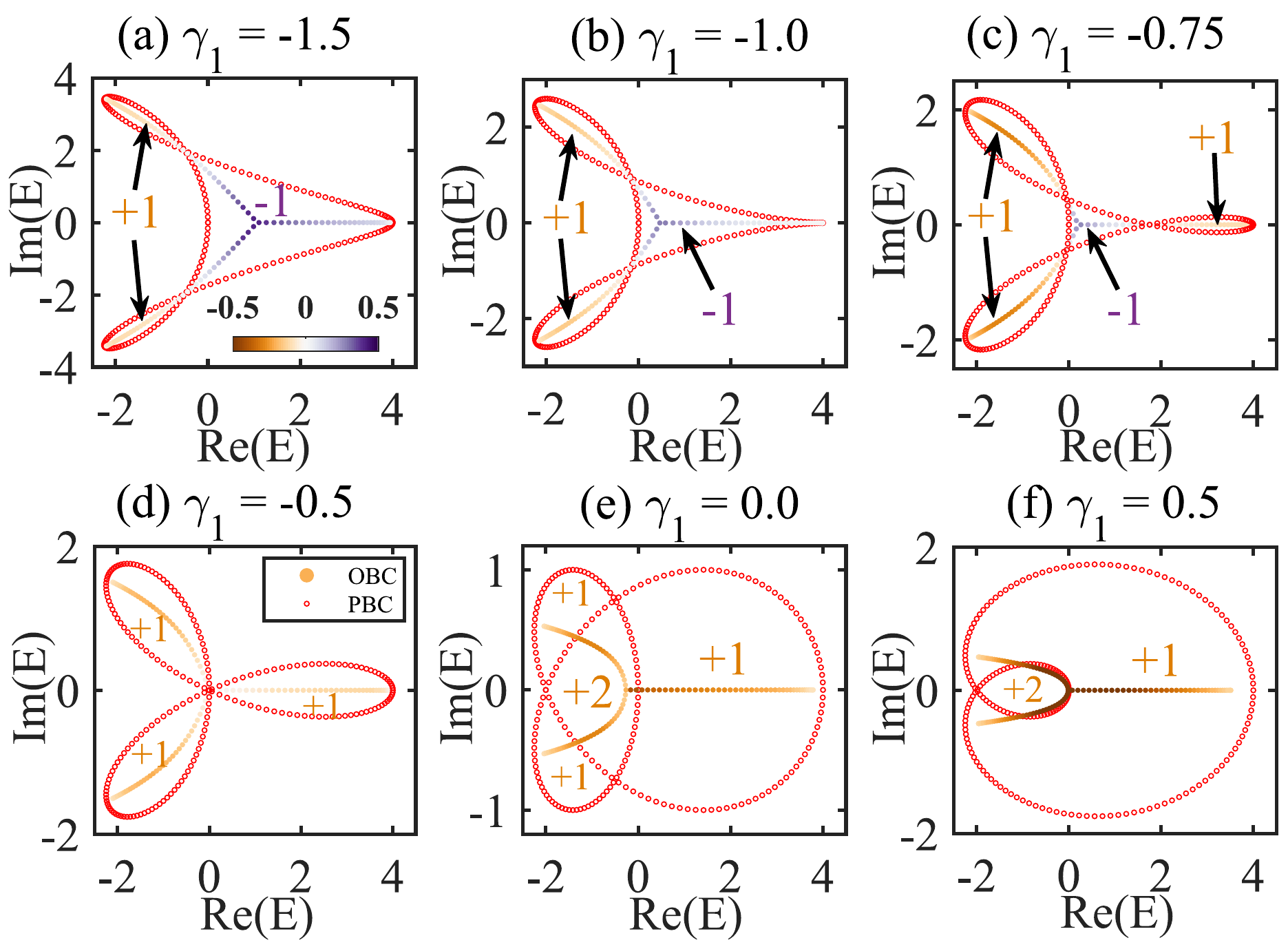}
	\caption{(Color online) Eigenenergy spectra of $H_2$ with $t_1=1$, $t_2=1$, and $\gamma_2=0.5$ with $\gamma_1$ varying from $-1.5$ to $0.5$ in (a)-(f). The solid dots correspond to the OBC spectra while the red empty dots are the PBC spectra. The numbers in the figures indicate the winding number when $E_B$ is located in the loop. The lattice size is $L=120$.}
	\label{figA2}
\end{figure}

To determine the NHSE edge, we can follow the method introduced in the previous subsection to calculate the self-intersecting points of the PBC spectra. Suppose that we have $k_1$, $k_2 \in [-\pi, \pi)$ and $k_1 \neq k_2$, such that $E_2(k_1)=E_2(k_2)$. Since $E_2(k)=2\left[t_1 \cos(k) + t_2 \cos(2k) \right] + 2i \left[\gamma_1 \sin(k) + \gamma_2 \sin(2k) \right]$, we have
\begin{equation}\label{k1k2}
	\begin{cases}
		t_1 \cos(k_1) + t_2 \cos(2k_1) = t_1 \cos(k_2) + t_2 \cos(2k_2), \\
		\gamma_1 \sin(k_1) + \gamma_2 \sin(2k_1) = \gamma_1 \sin(k_2) + \gamma_2 \sin(2k_2).
	\end{cases}
\end{equation}
By solving the above equations, we can obtain the self-intersecting points in the PBC spectra, and thus determine the skin effect edge in the OBC spectra. We first check the equation $t_1 \cos(k_1) + t_2 \cos(2k_1) = t_1 \cos(k_2) + t_2 \cos(2k_2)$. If $\cos(k_1) =\cos(k_2)$, then we have 
\begin{equation}
	t_1 \cos(k_1) =t_1 \cos(k_2) \quad and \quad t_2 \cos(2k_1)=t_2 \cos(2k_2) \Longrightarrow k_1 = -k_2.
\end{equation}
Substituting this into the other equation, we get 
\begin{equation}
	\gamma_1 \sin(k_1) + \gamma_2 \sin(2k_1) = \gamma_1 \sin(k_2) + \gamma_2 \sin(2k_2) = -\gamma_1 \sin(k_1) - \gamma_2 \sin(2k_1),
\end{equation}
which further leads to
\begin{equation}
	\sin(k_1)\left[\gamma_1 + 2\gamma_2 \cos(k_1)\right] = 0.
\end{equation}
If $\sin(k_1)=0$, then $k_1=\left\{ 0,-\pi \right\}$, and $k_2=-k_1=\left\{ 0,\pi \right\}$, this contradicts with the premise that $k_1$, $k_2 \in [-\pi, \pi)$ and $k_1 \neq k_2$. So if $\sin(k_1) \neq 0$, we have 
\begin{equation}
	\left[\gamma_1 + 2\gamma_2 \cos(k_1)\right] = 0 \Longrightarrow \cos(k_1) = -\frac{\gamma_1}{2\gamma_2}.
\end{equation}
Notice that this requires that $|\gamma_1/2\gamma_2|\leq 1$, i.e., $|\gamma_1|\leq |2\gamma_2|$. Substituting $\cos(k_1)$ back into $E_2(k)$, we obtain the self-intersecting point as 
\begin{equation}\label{SI1A}
	\text{Re}(E_2^{SI1}) = -\frac{\gamma_1}{\gamma_2}t_1 + \left( \frac{\gamma_1^2}{\gamma_2^2}-2 \right) t_2; \qquad \text{Im}(E_2^{SI1}) = 0.
\end{equation}
This gives us the real-valued NHSE edge in the spectrum (i.e., the red dot-dashed line in the spectrum). 

On the other hand, if $\cos(k_1) \neq \cos(k_2)$, then we can rewrite Eq.~(\ref{k1k2}) as 
\begin{equation}
	\begin{cases}
		\cos(k_1) + \cos(k_2) = -\frac{t_1}{2t_2} \\
		\gamma_1 \left[\sin(k_1) - \sin(k_2) \right] = \gamma_2 \left[\sin(2k_2) - \sin(2k_1) \right]
	\end{cases}
	\Longrightarrow
	\begin{cases}
		2\cos(\frac{k_1+k_2}{2}) \cos(\frac{k_1-k_2}{2}) = -\frac{t_1}{2t_2} \\
		2\gamma_1 \cos(\frac{k_1+k_2}{2}) \sin(\frac{k_1-k_2}{2}) = 2\gamma_2 \cos(k_2+k_1) \sin(k_2-k_1),
	\end{cases}
\end{equation}
which further leads to
\begin{equation}
	\begin{cases}
		\cos(\frac{k_1-k_2}{2}) = -\frac{t_1}{4t_2}/\cos(\frac{k_1+k_2}{2}) \\
		\gamma_1 \cos(\frac{k_1+k_2}{2}) = -2\gamma_2 \cos(k_1+k_2) \cos(\frac{k_1-k_2}{2}).
	\end{cases}
\end{equation}
Since $\cos(k_1+k_2)=2\cos^2(\frac{k_1+k_2}{2})-1$, by combining the above two equations, we can get
\begin{equation}
	\begin{split}
		& \cos^2 \left( \frac{k_1+k_2}{2} \right) = \frac{\frac{1}{2}t_1 \gamma_2}{t_1 \gamma_2 - t_2 \gamma_1}, \quad
		\sin^2 \left( \frac{k_1+k_2}{2} \right) = \frac{\frac{1}{2}t_1 \gamma_2-t_2 \gamma_1}{t_1 \gamma_2 - t_2 \gamma_1},\quad
		\cos \left( k_1+k_2 \right) = \frac{t_2 \gamma_1}{t_1 \gamma_2 - t_2 \gamma_1}, \\
		& \cos \left( \frac{k_1+k_2}{2} \right) = \pm \sqrt{ \frac{\frac{1}{2}t_1 \gamma_2}{t_1 \gamma_2 - t_2 \gamma_1} }, \quad
		\sin \left( \frac{k_1+k_2}{2} \right) = \pm \sqrt{ \frac{\frac{1}{2}t_1 \gamma_2-t_2 \gamma_1}{t_1 \gamma_2 - t_2 \gamma_1} }, \quad
		\sin(k_1+k_2) = \pm \frac{\sqrt{t_1^2 \gamma_2^2-2t_1 t_2 \gamma_1 \gamma_2}}{t_1\gamma_2-t_2\gamma_1}.
	\end{split}
\end{equation}
Then we have
\begin{equation}
	\cos \left(\frac{k_1-k_2}{2} \right) = -\frac{t_1}{4t_2}/\cos \left( \frac{k_1+k_2}{2} \right) = \mp \frac{t_1}{4 t_2} \sqrt{\frac{t_1 \gamma_2 -t_2 \gamma_1}{\frac{1}{2}t_1 \gamma_2}}, \quad
	\cos (k_1-k_2) = \frac{t_1(t_1 \gamma_2 - t_2 \gamma_1)}{4 t_2^2 \gamma_2} -1.
\end{equation}
With these trigonometric functions at hand, we can now calculate the self-intersecting points in the PBC spectra. From Eqs.~(\ref{E2k}) and (\ref{k1k2}), we know that the real part of the self-intersecting points $E_2^{SI}$ is
\begin{equation}
	\begin{split}
		\text{Re}(E_2^{SI}) &= t_1 \left[ \cos(k_1) + \cos(k_2) \right] + t_2 \left[ \cos(2k_1) + \cos(2k_2) \right] = -\frac{t_1^2}{2t_2} + 2t_2 \cos (k_1+k_2) \cos (k_1-k_2) \\
		&= -\frac{t_1^2}{2t_2} + 2t_2 \frac{t_2 \gamma_1}{t_1 \gamma_2 - t_2 \gamma_1} \left[ \frac{t_1(t_1 \gamma_2 - t_2 \gamma_1)}{4 t_2^2 \gamma_2} -1 \right] = -\frac{t_1^2}{2t_2} + \frac{t_1\gamma_1}{2\gamma_2} - \frac{2t_2^2 \gamma_1}{t_1 \gamma_2 - t_2 \gamma_1}.
	\end{split}
\end{equation}
Similarly, the imaginary part of $E_2^{SI}$ can be calculated as
\begin{equation}
	\begin{split}
		\text{Im}(E_2^{SI}) &= \gamma_1 \left[\sin(k_1) + \sin(k_2) \right] + \gamma_2 \left[\sin(2k_1) + \sin(2k_2) \right] \\
		&= 2 \gamma_1 \sin \left( \frac{k_1+k_2}{2} \right) \cos \left(\frac{k_1-k_2}{2} \right) + 2\gamma_2 \sin(k_1+k_2) \cos (k_1-k_2) \\
		&= \pm \frac{t_1 \gamma_1}{2 t_2} \sqrt{\frac{\frac{1}{2}t_1\gamma_2-t_2\gamma_1}{\frac{1}{2}t_1\gamma_2}} \mp 2\gamma_2 \frac{\sqrt{t_1^2 \gamma_2^2-2t_1 t_2 \gamma_1 \gamma_2}}{t_1\gamma_2-t_2\gamma_1} \left[   \frac{t_1(t_1 \gamma_2 - t_2 \gamma_1)}{4 t_2^2 \gamma_2} -1 \right] \\
		&= \sqrt{t_1^2 \gamma_2^2 - 2t_1 t_2 \gamma_1 \gamma_2} \left[ \pm \frac{\gamma_1}{2t_2 \gamma_2} \mp \left( \frac{t_1}{2t_2^2} - \frac{2\gamma_2}{t_1\gamma_2 - t_2\gamma_1} \right)\right]
	\end{split}.
\end{equation}
So, we get the other two self-intersecting points in the PBC spectrum as
\begin{equation}\label{SI23A}
	\begin{split}
		&\text{Re}(E_2^{SI2}) = -\frac{t_1^2}{2t_2} + \frac{t_1\gamma_1}{2\gamma_2} - \frac{2t_2^2 \gamma_1}{t_1 \gamma_2 - t_2 \gamma_1}; \quad \text{Im}(E_2^{SI2}) = \sqrt{t_1^2 \gamma_2^2 - 2t_1 t_2 \gamma_1 \gamma_2} \left[ \frac{\gamma_1}{2t_2 \gamma_2} - \left( \frac{t_1}{2t_2^2} - \frac{2\gamma_2}{t_1\gamma_2 - t_2\gamma_1} \right)\right]; \\
		&\text{Re}(E_2^{SI3}) = -\frac{t_1^2}{2t_2} + \frac{t_1\gamma_1}{2\gamma_2} - \frac{2t_2^2 \gamma_1}{t_1 \gamma_2 - t_2 \gamma_1}; \quad \text{Im}(E_2^{SI3}) = \sqrt{t_1^2 \gamma_2^2 - 2t_1 t_2 \gamma_1 \gamma_2} \left[ -\frac{\gamma_1}{2t_2 \gamma_2} + \left( \frac{t_1}{2t_2^2} - \frac{2\gamma_2}{t_1\gamma_2 - t_2\gamma_1} \right)\right].
	\end{split}
\end{equation}
Thus we have obtained the self-intersecting points in the systems with $t_2 \neq 0$. The NHSE edges in the OBC spectra are determined by these points, as shown in Fig.~\ref{figA1}. 

As the system parameters vary, the skin effect edges will cross each other. The crossing point corresponds to the merging point of the self-intersections in the PBC spectrum, which is obtained by solving the following equation
\begin{equation}
	E_2^{SI1}=E_2^{SI2}=E_2^{SI3}.
\end{equation}
Then we get the following equation for the real part of the $E_2^{SI}$:
\begin{equation}
	-\frac{\gamma_1}{\gamma_2}t_1 + \left( \frac{\gamma_1^2}{\gamma_2^2}-2 \right) t_2 = -\frac{t_1^2}{2t_2} + \frac{t_1\gamma_1}{2\gamma_2} - \frac{2t_2^2 \gamma_1}{t_1 \gamma_2 - t_2 \gamma_1}.
\end{equation}
Defining $\gamma=\frac{\gamma_1}{\gamma_2}$, the above equation can be rewritten as
\begin{equation}
	(2\gamma^3-8\gamma)t_2^3-(5\gamma^2-4)t_1 t_2^2 + 4\gamma t_1^2 t_2 - t_1^3 = 0.
\end{equation}

Taking the system with $t_1=t_2=1$ as an example, after substituting $t_1=t_2=1$ into it, we obtain the following cubic equation
\begin{equation}
	2\gamma^3-5\gamma^2-4\gamma+3=0.
\end{equation}
Comparing with the standard formula of the cubic equation $ax^3+bx^2+cx+d=0$, we have $a=2,b=-5,c=-4$, and $d=3$. By defining $p=\frac{3ac-b^2}{3a^2}$ and $q=\frac{27a^2d-9abc+2b^3}{27a^3}$, the equation discriminant is
\begin{equation}
	\Delta = \frac{q^2}{4} + \frac{p^3}{27} > 0.
\end{equation}
Then the cubic equation has three roots, one real and two complex. Since $\gamma_1$ and $\gamma_2$ are real numbers, we only take the real root, which turns out to be 
\begin{equation}
	\gamma=-1 \longrightarrow \gamma_1 = -\gamma_2.
\end{equation}
The corresponding imaginary parts for the edges are all zero at this point. So this is the critical parameter when the the edges merge into one single point, consistent with the numerical results shown in Fig.~\ref{figA1} where the skin effect edges cross at $\gamma_1=-0.5$ for the system with $\gamma_2=0.5$.

\subsection{Non-Hermitian skin effect edge in systems with $\mathbf{r_d=3}$ and $\mathbf{r_d=4}$}
Here we further investigate the non-Hermitian lattices with third- and fourth-neighboring hopping. The general model Hamiltonian for the systems with hopping terms up to the third-neighboring sites is described as
\begin{equation}\label{H3}
	H_3 = \sum_{j=1}^{L-3} \left[ (t_1-\gamma_1)c_{j+1}^\dagger c_{j} + (t_2-\gamma_2) c_{j+2}^\dagger c_{j} + (t_3-\gamma_3) c_{j+3}^\dagger c_{j} + (t_1+\gamma_1) c_{j}^\dagger c_{j+1} + (t_2+\gamma_2) c_{j}^\dagger c_{j+2} + (t_3+\gamma_3) c_{j}^\dagger c_{j+3} \right],
\end{equation}
where the parameters $t_i$ and $\gamma_i$ $(i=1,2,3)$ are set to be real numbers. To check the influence of the nonreciprocal hopping in the third-neighboring hopping terms, we can set $t_2=t_3=0$ and $\gamma_1=\gamma_2=0$, and obtain the following simplified Hamiltonian
\begin{equation}
	H_3^\prime = \sum_{j=1}^{L-3} \left[ t_1 c_{j+1}^\dagger c_{j} + t_1 c_{j}^\dagger c_{j+1} -\gamma_3 c_{j+3}^\dagger c_{j} +  \gamma_3 c_{j}^\dagger c_{j+3} \right],
\end{equation}
The OBC spectra are obtained by diagonalizing the above Hamiltonian and the PBC spectra can be obtained by transforming the Hamiltonian into the momentum space and get
\begin{equation}
	E_3^\prime(k) = 2t_1 \cos(k) + 2i \gamma_3 \sin(3k).
\end{equation}
To determine the critical energy, we can calculate the self-intersecting points in the PBC spectrum. Following the method introduced in the above sections, we can suppose that there are $k_1$, $k_2 \in [-\pi, \pi)$ and $k_1 \neq k_2$, such that $E_3^\prime(k_1)=E_3^\prime(k_2)$. Then we have
\begin{equation}
	\begin{cases}
		t_1 \cos(k_1) = t_1 \cos(k_2) \\
		\gamma_3 \sin(3k_1) = \gamma_3 \sin(3k_2)
	\end{cases}
	\Longrightarrow
	\begin{cases}
		\cos(k_1) = \cos(k_2) \\
		3\sin(k_1) - 4\sin^3(k_1) = 3\sin(k_2) - 4\sin^3(k_2),
	\end{cases}
\end{equation} 
which further leads to
\begin{equation}
	\begin{cases}
		k_1 = -k_2 \\
		3 \left[ \sin(k_1) - \sin(k_2) \right] = 4 \left[\sin^3(k_1) - \sin^3(k_2) \right]
	\end{cases}
	\Longrightarrow
	\begin{cases}
		k_1 = -k_2 \\
		\sin^2(k_1) + \sin(k_1) \sin(k_2) + \sin^2(k_2) = \frac{3}{4}.
	\end{cases}
\end{equation}
Solving the equations gives us $k_1=\pm \frac{\pi}{3}$. Substituting $k_1$ into $E_3^\prime(k)$, we can obtain two self-intersecting points as
\begin{align}
	\text{Re}\left(E_3^{SI1}\right) = +t_1, \quad \text{Im}\left(E_3^{SI1}\right) = 0; \\ \notag
	\text{Re}\left(E_3^{SI2}\right) = -t_1, \quad \text{Im}\left(E_3^{SI2}\right) = 0.
\end{align}

Similarly, the model Hamiltonian for the systems with hopping terms up to the fourth-neighboring sites is
\begin{align}\label{H4}
	H_4 = \sum_{j=1}^{L-4} & \left[ (t_1-\gamma_1)c_{j+1}^\dagger c_{j} + (t_2-\gamma_2) c_{j+2}^\dagger c_{j} + (t_3-\gamma_3) c_{j+3}^\dagger c_{j} + (t_4-\gamma_4) c_{j+4}^\dagger c_{j} \right. \\ \notag
	& \left. + (t_1+\gamma_1) c_{j}^\dagger c_{j+1} + (t_2+\gamma_2) c_{j}^\dagger c_{j+2} + (t_3+\gamma_3) c_{j}^\dagger c_{j+3} +(t_4+\gamma_4) c_{j}^\dagger c_{j+4} \right].
\end{align}
We can also check the simplified Hamiltonian which accounts for the nonreciprocal hopping in the fourth-neighboring hopping terms and get
\begin{equation}
	H_4^\prime = \sum_{j=1}^{L-4} \left[ t_1 c_{j+1}^\dagger c_{j} + t_1 c_{j}^\dagger c_{j+1} -\gamma_4 c_{j+4}^\dagger c_{j} +  \gamma_4 c_{j}^\dagger c_{j+4} \right].
\end{equation}
The corresponding PBC spectra are 
\begin{equation}
	E_4^\prime(k) = 2t_1 \cos(k) + 2i \gamma_4 \sin(4k).
\end{equation}
Again we can determine the skin effect edge by calculating the self-intersecting points. Suppose that there are $k_1$, $k_2 \in [-\pi, \pi)$ and $k_1 \neq k_2$, such that $E_4^\prime(k_1)=E_4^\prime(k_2)$. Then we have
\begin{equation}
	\begin{cases}
		t_1 \cos(k_1) = t_1 \cos(k_2) \\
		\gamma_4 \sin(4k_1) = \gamma_4 \sin(4k_2)
	\end{cases}
	\Longrightarrow
	\begin{cases}
		\cos(k_1) = \cos(k_2) \\
		\cos(k_1) \sin(k_1) \left[2 \sin^2(k_1)-1 \right] = \cos(k_2) \sin(k_2) \left[2 \sin^2(k_2)-1 \right],
	\end{cases}
\end{equation}
which further leads to 
\begin{equation}
	\begin{cases}
		k_1 = -k_2 \\
		\sin(k_1) \left[2 \sin^2(k_1)-1 \right] = \sin(k_2) \left[2 \sin^2(k_2)-1 \right].
	\end{cases}
\end{equation}
If $k_1=\pm\pi/2$, the above two conditions can be satisfied, and we can obtain one self-intersecting point as
\begin{equation}
	\text{Re}\left(E_4^{SI1}\right) = 0, \quad \text{Im}\left(E_4^{SI1}\right) = 0.
\end{equation}
If $k \neq \pm\pi/2$, then from these two equations we can get $\sin(k_1)= \pm \frac{\sqrt{2}}{2}$, and we have $k_1 = \{\pi/4, 3\pi/4\}$. Substituting $k_1$ into $E_4^\prime(k)$, we get another two self-intersecting points in the PBC spectrum:
\begin{align}
	& \text{Re}\left(E_4^{SI2}\right) = +\sqrt{2}t_1, \quad \text{Im}\left(E_4^{SI2}\right) = 0.\\ \notag
	& \text{Re}\left(E_4^{SI3}\right) = -\sqrt{2}t_1, \quad \text{Im}\left(E_4^{SI3}\right) = 0.
\end{align}
Thus we obtain the three self-intersecting points in the PBC spectrum and determine the three NHSE edges in the OBC spectrum, which are consistent with the numerical results shown in Fig.~5 in the main text.

For the general cases with nonreciprocal hopping up to the third- and fourth-neighboring sites, the same method can be utilized to analyze the NHSE edges.

\end{widetext}


\begin{thebibliography}{}
\bibitem{Cao2015RMP}{H. Cao and J. Wiersig, \href{https://doi.org/10.1103/RevModPhys.87.61}{Rev. Mod. Phys. \textbf{87,} 61 (2015).}}

\bibitem{Konotop2016RMP}{V. V. Konotop, J. Yang, and D. A. Zezyulin, \href{https://doi.org/10.1103/RevModPhys.88.035002}{Rev. Mod. Phys. \textbf{88,} 035002 (2016).}}

\bibitem{Ganainy2018NatPhy}{R. El-Ganainy, K. G. Makris, M. Khajavikhan, Z. H. Musslimani, S. Rotter, and D. N. Christodoulides, \href{https://doi.org/10.1038/nphys4323}{Nat. Phys. \textbf{14,} 11 (2018).}}

\bibitem{Ashida2020AiP}{Y. Ashida, Z. Gong, and M. Ueda, \href{https://doi.org/10.1080/00018732.2021.1876991}{Advances in Physics \textbf{69,} 249 (2020).}}

\bibitem{Bergholtz2021RMP}{E. J. Bergholtz, J. C. Budich, and F. K. Kunst, \href{https://doi.org/10.1103/RevModPhys.93.015005}{Rev. Mod. Phys. \textbf{93,} 015005 (2021).}}	

\bibitem{Makris2008PRL}{K. G. Makris, R. El-Ganainy, D. N. Christodoulides, and Z. H. Musslimani, \href{https://doi.org/10.1103/PhysRevLett.100.103904}{Phys. Rev. Lett. \textbf{100,} 103904 (2008).}}

\bibitem{Klaiman2008PRL}{S. Klaiman, U. Günther, and N. Moiseyev, \href{https://doi.org/10.1103/PhysRevLett.101.080402}{Phys. Rev. Lett. \textbf{101,} 080402 (2008).}}

\bibitem{Guo2009PRL}{A. Guo, G. J. Salamo, D. Duchesne, R. Morandotti, M. VolatierRavat, V. Aimez, G. A. Siviloglou, and D. N. Christodoulides, \href{https://doi.org/10.1103/PhysRevLett.103.093902}{Phys. Rev. Lett. \textbf{103,} 093902 (2009).}}

\bibitem{Ruter2010NatPhys}{C. E. Rüter, K. G. Makris, R. El-Ganainy, D. N. Christodoulides, M. Segev, and D. Kip, \href{https://doi.org/10.1038/nphys1515}{Nat. Phys. \textbf{6,} 192 (2010).}}

\bibitem{Lin2011PRL}{Z. Lin, H. Ramezani, T. Eichelkraut, T. Kottos, H. Cao, and D. N. Christodoulides, \href{https://doi.org/10.1103/PhysRevLett.106.213901}{Phys. Rev. Lett. \textbf{106,} 213901 (2011).}}

\bibitem{Regensburger2012Nat}{A. Regensburger, C. Bersch, M.-A. Miri, G. Onishchukov, D. N. Christodoulides, and U. Peschel, \href{https://doi.org/10.1038/nature11298}{Nature (London) \textbf{488,} 167 (2012).}}

\bibitem{Feng2013NatMat}{L. Feng, Y.-L. Xu, W. S. Fegadolli, M.-H. Lu, J. E. B. Oliveira, V. R. Almeida, Y.-F. Chen, and A. Scherer, \href{https://doi.org/10.1038/nmat3495}{Nat. Mater. \textbf{12,} 108 (2013).}}

\bibitem{Peng2014NatPhys}{B. Peng, S. K. Özdemir, F. Lei, F. Monifi, M. Gianfreda, G. L. Long, S. Fan, F. Nori, C. M. Bender, and L. Yang, \href{https://doi.org/10.1038/nphys2927}{Nat. Phys. \textbf{10,} 394 (2014).}}

\bibitem{Wiersig2014PRL}{J. Wiersig, \href{https://doi.org/10.1103/PhysRevLett.112.203901}{Phys. Rev. Lett. \textbf{112,} 203901 (2014).}}

\bibitem{Hodaei2017Nat}{H. Hodaei, A. U. Hassan, S. Wittek, H. Garcia-Gracia, R. El-Ganainy, D. N. Christodoulides, and M. Khajavikhan, \href{https://doi.org/10.1038/nature23280}{Nature (London) \textbf{548,} 187 (2017).}}

\bibitem{Chen2017Nat}{W. Chen, ¸ S. K. Özdemir, G. Zhao, J. Wiersig, and L. Yang, \href{https://doi.org/10.1038/nature23281}{Nature (London) \textbf{548,} 192 (2017).}}

\bibitem{Brody2012PRL}{D. C. Brody and E.-M. Graefe, \href{https://doi.org/10.1103/PhysRevLett.109.230405}{Phys. Rev. Lett. \textbf{109,} 230405 (2012).}}

\bibitem{Lee2014PRX}{T. E. Lee and C.-K. Chan, \href{https://doi.org/10.1103/PhysRevX.4.041001}{Phys. Rev. X \textbf{4,} 041001 (2014).}}

\bibitem{Li2019NatCom}{J. Li, A. K. Harter, J. Liu, L. de Melo, Y. N. Joglekar, and L. Luo, \href{https://doi.org/10.1038/s41467-019-08596-1}{Nat. Commun. \textbf{10,} 855 (2019).}}

\bibitem{Kawabata2017PRL}{K. Kawabata, Y. Ashida, and M. Ueda, \href{https://doi.org/10.1103/PhysRevLett.119.190401}{Phys. Rev. Lett. \textbf{119,} 190401 (2017).}}

\bibitem{Hamazaki2019PRL}{R. Hamazaki, K. Kawabata, and M. Ueda, \href{https://doi.org/10.1103/PhysRevLett.123.090603}{Phys. Rev. Lett. \textbf{123,} 090603 (2019).}}

\bibitem{Xiao2019PRL}{L. Xiao, K. Wang, X. Zhan, Z. Bian, K. Kawabata, M. Ueda, W. Yi, and P. Xue, \href{https://doi.org/10.1103/PhysRevLett.123.230401}{Phys. Rev. Lett. \textbf{123,} 230401 (2019).}}

\bibitem{Wu2019Science}{Y. Wu, W. Liu, J. Geng, X. Song, X. Ye, C.-K. Duan, X. Rong, and J. Du, \href{https://doi.org/10.1126/science.aaw8205}{Science \textbf{364,} 878 (2019).}}

\bibitem{Yamamoto2019PRL}{K. Yamamoto, M. Nakagawa, K. Adachi, K. Takasan, M. Ueda, and N. Kawakami, \href{https://doi.org/10.1103/PhysRevLett.123.123601}{Phys. Rev. Lett. \textbf{123,} 123601 (2019).}}

\bibitem{Naghiloo2019NatPhys}{M. Naghiloo, N. Abbasi, Y. N. Joglekar, and K. W. Murch, \href{https://doi.org/10.1038/s41567-019-0652-z}{Nat. Phys. \textbf{15,} 1232 (2019).}}

\bibitem{Matsumoto2020PRL}{N. Matsumoto, K. Kawabata, Y. Ashida, S. Furukawa, and M. Ueda, \href{https://doi.org/10.1103/PhysRevLett.125.260601}{Phys. Rev. Lett. \textbf{125,} 260601 (2020).}}

\bibitem{Bender1998PRL}{C. M. Bender and S. Boettcher, \href{https://doi.org/10.1103/PhysRevLett.80.5243}{Phys. Rev. Lett. \textbf{80,} 5243 (1998).}}

\bibitem{Bender2002PRL}{C. M. Bender, D. C. Brody, and H. F. Jones, \href{https://doi.org/10.1103/PhysRevLett.89.270401}{Phys. Rev. Lett. \textbf{89,} 270401 (2002).}}

\bibitem{Bender2007RPP}{C. M. Bender, \href{https://doi.org/10.1088/0034-4885/70/6/R03}{Rep. Prog. Phys. \textbf{70,} 947 (2007).}}

\bibitem{Mostafazadeh2002JMP}{A. Mostafazadeh, \href{https://doi.org/10.1063/1.1418246}{J. Math. Phys. \textbf{43,} 205 (2002);} \href{ https://doi.org/10.1063/1.1461427}{\emph{ibid.} \textbf{43,} 2814 (2002);} \href{https://doi.org/10.1063/1.1489072}{\emph{ibid.} \textbf{43,} 3944 (2002).}}

\bibitem{Mostafazadeh2010IJMMP}{A. Mostafazadeh, \href{https://doi.org/10.1142/S0219887810004816}{Int. J. Geom. Meth. Mod. Phys. \textbf{7,} 1191 (2010).}}

\bibitem{Moiseyev2011Book}{N. Moiseyev, \emph{Non-Hermitian Quantum Mechanics} (Cambridge University Press, Cambridge, UK, 2011).}

\bibitem{Zeng2020PRB1}{Q.-B. Zeng, Y.-B. Yang, and R. L\"u, \href{https://doi.org/10.1103/PhysRevB.101.125418}{Phys. Rev. B \textbf{101,} 125418 (2020).}}

\bibitem{Kawabata2020PRR}{K. Kawabata and M. Sato, \href{https://doi.org/10.1103/PhysRevResearch.2.033391}{Phys. Rev. Research \textbf{2,} 033391 (2020).}}

\bibitem{Zeng2021NJP}{Q.-B. Zeng and Rong L\"u, \href{https://doi.org/10.1088/1367-2630/ac61d0}{New J. Phys. \textbf{24,} 043023 (2022).}}

\bibitem{Yao2018PRL1}{S. Yao and Z. Wang, \href{https://doi.org/10.1103/PhysRevLett.121.086803}{Phys. Rev. Lett. \textbf{121,} 086803 (2018).}}

\bibitem{Yao2018PRL2}{S. Yao, F. Song, and Z. Wang, \href{https://doi.org/10.1103/PhysRevLett.121.136802}{Phys. Rev. Lett. \textbf{121,} 136802 (2018).}}

\bibitem{Alvarez2018PRB}{V. M. Martinez Alvarez, J. E. Barrios Vargas, and L. E. F. Foa Torres, \href{https://doi.org/10.1103/PhysRevB.97.121401}{Phys. Rev. B \textbf{97,} 121401(R) (2018).}}

\bibitem{Alvarez2018EPJ}{V. M. Martinez Alvarez, J. E. Barrios Vargas, M. Berdakin, and L. E. F. Foa Torres, \href{https://doi.org/10.1140/epjst/e2018-800091-5}{Eur. Phys. J. Spec. Top. \textbf{227,} 1295 (2018).}}

\bibitem{Lee2019PRB}{C. H. Lee and R. Thomale, \href{https://journals.aps.org/prb/abstract/10.1103/PhysRevB.99.201103}{Phys. Rev. B \textbf{99,} 201103(R) (2019).}}

\bibitem{Zhou2019PRB}{H. Zhou and J. Y. Lee, \href{https://doi.org/10.1103/PhysRevB.99.235112}{Phys. Rev. B \textbf{99,} 235112 (2019).}}

\bibitem{Kawabata2019PRX}{K. Kawabata, K. Shiozaki, M. Ueda, and M. Sato, \href{https://doi.org/10.1103/PhysRevX.9.041015}{Phys. Rev. X \textbf{9,} 041015 (2019).}}

\bibitem{Song2019PRL}{F. Song, S. Yao, and Z. Wang, \href{https://doi.org/10.1103/PhysRevLett.123.170401}{Phys. Rev. Lett. \textbf{123,} 170401 (2019).}}

\bibitem{Okuma2020PRB}{N. Okuma and Masatoshi Sato, \href{https://doi.org/10.1103/PhysRevB.102.014203}{Phys. Rev. B \textbf{102,} 014203 (2020).}}

\bibitem{Xiao2020NatPhys}{L. Xiao, T. Deng, K. Wang, G. Zhu, Z. Wang, W. Yi, and P. Xue, \href{https://doi.org/10.1038/s41567-020-0836-6}{Nat. Phys. \textbf{16,} 761 (2020).}}

\bibitem{Yoshida2020PRR}{T. Yoshida, T. Mizoguchi, and Y. Hatsugai, \href{https://doi.org/10.1103/PhysRevResearch.2.022062}{Phys. Rev. Research \textbf{2,} 022062(R) (2020).}}

\bibitem{Longhi2019PRR}{S. Longhi, \href{https://doi.org/10.1103/PhysRevResearch.1.023013}{Phys. Rev. Research \textbf{1,} 023013 (2019).}}

\bibitem{Yi2020PRL}{Y. Yi and Z. Yang, \href{https://doi.org/10.1103/PhysRevLett.125.186802}{Phys. Rev. Lett. \textbf{125,} 186802 (2020).}}

\bibitem{Claes2021PRB}{J. Claes and T. L. Hughes, \href{https://doi.org/10.1103/PhysRevB.103.L140201}{Phys. Rev. B \textbf{103,} L140201 (2021).}}

\bibitem{Haga2021PRL}{T. Haga, M. Nakagawa, R. Hamazaki, and M. Ueda, \href{https://doi.org/10.1103/PhysRevLett.127.070402}{Phys. Rev. Lett. \textbf{127,} 070402 (2021).}}

\bibitem{Zeng2022PRA}{Q.-B. Zeng and R. L\"u, \href{https://doi.org/10.1103/PhysRevA.105.042211}{Phys. Rev. A \textbf{105,} 042211 (2022).}} 

\bibitem{Zeng2022PRB}{Q.-B. Zeng and R. L\"u, \href{https://doi.org/10.1103/PhysRevB.105.245407}{Phys. Rev. B \textbf{105,} 245407 (2022).}}

\bibitem{Kunst2018PRL}{F. K. Kunst, E. Edvardsson, J. C. Budich, and E. J. Bergholtz, \href{https://doi.org/10.1103/PhysRevLett.121.026808}{Phys. Rev. Lett. \textbf{121,} 026808 (2018).}}

\bibitem{Jin2019PRB}{L. Jin and Z. Song, \href{https://doi.org/10.1103/PhysRevB.99.081103}{Phys. Rev. B \textbf{99,} 081103(R) (2019).}}

\bibitem{Yokomizo2019PRL}{K. Yokomizo and S. Murakami, \href{https://doi.org/10.1103/PhysRevLett.123.066404}{Phys. Rev. Lett. \textbf{123,} 066404 (2019).}}

\bibitem{Herviou2019PRA}{L. Herviou, J. H. Bardarson, and N. Regnault, \href{https://doi.org/10.1103/PhysRevA.99.052118}{Phys. Rev. A \textbf{99,} 052118 (2019).}}

\bibitem{Zeng2020PRB}{Q.-B. Zeng, Y.-B. Yang, and Y. Xu, \href{https://doi.org/10.1103/PhysRevB.101.020201}{Phys. Rev. B 101, 020201(R) (2020).}}

\bibitem{Borgnia2020PRL}{D. S. Borgnia, A. J. Kruchkov, and R.-J. Slager, \href{https://doi.org/10.1103/PhysRevLett.124.056802}{Phys. Rev. Lett. \textbf{124,} 056802 (2020).}}

\bibitem{Yang2020PRL2}{Z. Yang, K. Zhang, C. Fang, and J. Hu, \href{https://doi.org/10.1103/PhysRevLett.125.226402}{Phys. Rev. Lett. \textbf{125,} 226402 (2020).}}

\bibitem{Zirnstein2021PRL}{H.-G. Zirnstein, G. Refael, and B. Rosenow, \href{https://doi.org/10.1103/PhysRevLett.126.216407}{Phys. Rev. Lett. \textbf{126,} 216407 (2021).}}

\bibitem{Zhang2022arxiv}{Z. Q. Zhang, H. Liu, H. Liu, H. Jiang, and X. C. Xie, \href{https://doi.org/10.48550/arXiv.2201.01577}{arXiv:2201.01577.}}

\bibitem{Xiong2018JPC}{Y. Xiong, \href{https://doi.org/10.1088/2399-6528/aab64a}{J. Phys. Commun. \textbf{2,} 035043 (2018).}}

\bibitem{Budich2020PRL}{J. C. Budich and E. J. Bergholtz, \href{https://doi.org/10.1103/PhysRevLett.125.180403}{Phys. Rev. Lett. \textbf{125,} 180403 (2020).}}

\bibitem{Koch2022PRR}{F. Koch and J. C. Budich, \href{https://doi.org/10.1103/PhysRevResearch.4.013113}{Phys. Rev. Research \textbf{4,} 013113 (2022).}}

\bibitem{Hatano1996PRL}{N. Hatano and D. R. Nelson, \href{https://doi.org/10.1103/PhysRevLett.77.570}{Phys. Rev. Lett. \textbf{77,} 570 (1996).}}

\bibitem{Shnerb1998PRL}{N. M. Shnerb and D. R. Nelson, \href{https://doi.org/10.1103/PhysRevLett.80.5172}{Phys. Rev. Lett. \textbf{80,} 5172 (1998).}}

\bibitem{Gong2018PRX}{Z. Gong, Y. Ashida, K. Kawabata, K. Takasan, S. Higashikawa, and M. Ueda, \href{https://doi.org/10.1103/PhysRevX.8.031079}{Phys. Rev. X \textbf{8,} 031079 (2018).}}

\bibitem{Jiang2019PRB}{H. Jiang, L.-J. Lang, C. Yang, S.-L. Zhu, and S. Chen, \href{https://doi.org/10.1103/PhysRevB.100.054301}{Phys. Rev. B \textbf{100,} 054301 (2019).}}

\bibitem{Zeng2020PRR}{Q.-B. Zeng and Y. Xu, \href{https://doi.org/10.1103/PhysRevResearch.2.033052}{Phys. Rev. Research \textbf{2,} 033052 (2020).}}

\bibitem{Liu2021PRB1}{Y. Liu, Y. Wang, X. J. Liu, Q. Zhou, and S. Chen, \href{https://doi.org/10.1103/PhysRevB.103.014203}{Phys. Rev. B \textbf{103,} 014203 (2021).}}

\bibitem{Liu2021PRB2}{Y. Liu, Q. Zhou, and S. Chen, \href{https://doi.org/10.1103/PhysRevB.104.024201}{Phys. Rev. B \textbf{104,} 024201 (2021).}}

\bibitem{Biddle2011PRB}{J. Biddle, D. J. Priour Jr., B. Wang, and S. Das Sarma, \href{https://doi.org/10.1103/PhysRevB.83.075105}{Phys. Rev. B \textbf{83,} 075105 (2011).}}

\bibitem{Okuma2020PRL}{N. Okuma, K. Kawabata, K. Shiozaki, and M. Sato, \href{https://doi.org/10.1103/PhysRevLett.124.086801}{Phys. Rev. Lett. \textbf{124,} 086801 (2020).}}

\bibitem{Zhang2020PRL}{K. Zhang, Z. Yang, and C. Fang, \href{https://doi.org/10.1103/PhysRevLett.125.126402}{Phys. Rev. Lett. \textbf{125,} 126402 (2020).}}

\end{thebibliography}
\end{document}